\documentclass[12pt]{iopart}
\usepackage{graphicx}
\usepackage{xspace}
\usepackage{amssymb}
\usepackage{soul}
\usepackage{bm}
\usepackage{tikz}
\usepackage{xcolor}

\renewcommand{\etal}{\emph{et al}.\xspace}
\newcommand{\uline}{\underline}

\newcommand{\kT}{k_B T}
\newcommand{\bG}{{\bm G}}
\newcommand{\bphi}{{\bm \phi}}
\newcommand{\bb}{{\bm b}}

\usepackage{hyperref}
\usepackage{url}
\definecolor{lightred}{rgb}{1,0.85,0.88}
\hypersetup{colorlinks,breaklinks,
            linkcolor=blue,urlcolor=blue,
            anchorcolor=blue,citecolor=blue}

\begin{document}

\title[Coarse-grained ionic liquids]
{Dynamical properties across different coarse-grained models for ionic liquids}

\author{
Joseph F.~Rudzinski$^1$, Sebastian Kloth,$^2$ Svenja W\"orner,$^1$ Tamisra Pal,$^2$ Kurt Kremer,$^1$ Tristan Bereau,$^{3,1}$ Michael Vogel$^2$
}

\address{$^1$ Max Planck Institute for Polymer Research, 55128 Mainz,
Germany}

\address{$^2$ Institute of Condensed Matter Physics, Technische
Universit\"at Darmstadt, Hochschulstr.~6, 64289 Darmstadt, Germany}

\address{$^3$ Van 't Hoff Institute for Molecular Sciences and
Informatics Institute, University of Amsterdam, Amsterdam 1098 XH, The
Netherlands}

\ead{\mailto{rudzinski@mpip-mainz.mpg.de}}

\begin{abstract}
    Room-temperature ionic liquids (RTILs) stand out among molecular liquids for their rich physicochemical characteristics, including structural and dynamic heterogeneity. The significance of electrostatic interactions in RTILs results in long characteristic length- and timescales, and has motivated the development of a number of coarse-grained (CG) simulation models. In this study, we aim to better understand the connection between certain CG parametrization strategies and the dynamical properties and transferability of the resulting models. We systematically compare five CG models: a model largely parametrized from experimental thermodynamic observables; a refinement of this model to increase its structural accuracy; and three models that reproduce a given set of structural distribution functions by construction, with varying intramolecular parametrizations and reference temperatures. All five CG models display limited structural transferability over temperature, and also result in various effective dynamical speedup factors, relative to a reference atomistic model. On the other hand, the structure-based CG models tend to result in more consistent cation-anion relative diffusion than the thermodynamic-based models, for a single thermodynamic state point. By linking short- and long-timescale dynamical behaviors, we demonstrate that the varying dynamical properties of the different coarse-grained models can be largely collapsed onto a single curve, which provides evidence for a route to constructing dynamically-consistent CG models of RTILs.
\end{abstract}
\submitto{\JPCM}
\maketitle

\section{Introduction}

Of the broad variety of molecular liquids, ionic liquids (ILs) stand
out for their rich physicochemical characteristics~\cite{Atkin_CR_15,
Fayer_JCP_18}. ILs are salts, with a melting point or glass-transition
temperature that can reach low temperatures---notably,
``room-temperature'' ILs (RTILs) are in the liquid state at ambient conditions.
RTILs are commonly composed of an organic cation and an inorganic anion. ILs
play an important role as a solvent in sustainable chemistry, with
applications including biomaterials and catalysis~\cite{armand2009ionic, Armand_11, Angell_EES_14, Watanabe_CR_17}.
Being conductive, ILs are also strong candidates in electrochemical
applications. 

The organic cations in ILs often consist of a polar ring group along with nonpolar side chains. 
The amphiphilic nature of these cations facilitates the
formation of nanoscale segregation. Nanoheterogeneous structures have
been investigated using both X-ray and neutron scattering~\cite{Russina_JPCB_07, Margulis_JPCB_10, Hardacre_JCP_10,
Russina_JPCL_12, Yamamuro_JCP_15}. Analysis of the structure factor
revealed structural inhomogeneity, together with complex heterogeneous dynamics, observed via dielectric spectroscopy~\cite{Rivera_JCP_07, Kremer_PCCP_09}. Nanoscale
segregation increases when the temperature is decreased toward the glass
transition, evidenced by shifts in the static structure factor~\cite{Russina_JPCL_12}. 
Low temperatures can also yield specific dynamical effects,
such as a breakdown of the Stokes-Einstein-Debye relation~\cite{Vogel_JCP_19_MB}. 
Despite a wealth of studies characterizing the properties of ILs, a clear link between the structural and
dynamical properties of these systems, especially close to the glassy regime, remains elusive.

Computer simulations have played a significant role in furthering our
understanding of ILs~\cite{Voth_JPCB_04, Padua_JPCB_06, wang2007understanding,
DelleSitte_FD_11, Lopes_JBCS_16}. A review by Maginn highlights
that interests in ILs coincided with the advent and development of
molecular simulations~\cite{Maginn2009}. Simulations have
provided invaluable insight into the structural, thermodynamic, and
dynamical aspects of ILs~\cite{Plate_CPC_10, Wang_JCP_16,
Kim_JCP_18, Sulpizi_JCP_18, Smiatek_JCP_18}. The dramatic breadth of
relevant length- and timescales has made excellent use of multiscale
modeling---from quantum mechanics to classical atomistic to
coarse-graining---to shed light on various aspects including
viscosity, interfacial behavior, and dynamic heterogeneity~\cite{Singer_JPCA_03, Popolo_JPCB_05, Balasubramanian_CPL_06,
wang2007understanding, lynden2007simulations, bhargava2008modelling, Vogel_JPCC_18}. 

Going up the multiscale-modeling ladder facilitates the study of
phenomena occurring at longer length- and timescales. 
The need for large systems and the
associated extensive timescales, due in no small part to the strong electrostatic
interactions in ILs, lend themselves to a coarse-grained
(CG) description of the system. By lumping together several atoms into
super-particles or beads, CG models not only decrease the number of particles
to be simulated, but also effectively smooth the underlying free-energy
landscape~\cite{Peter:2009hr,Riniker:2012qf,noid2013perspective}.

A variety of CG approaches have been previously applied to study ILs.
Wang and Voth used the force-matching-based multiscale coarse-graining method to
coarse-grain 1-$n$-alkyl-3-methylimidazolium ([C$_n$mim]$^+$) nitrate,
for both $n=2$ (ethyl) and $n=4$ (butyl)~\cite{WangVoth2006}. The
mapping involved a single bead for the imidazolium ring, one bead per
methylene group, as well as a separate bead for the terminal methyl group.
Explicit Coulomb interactions were used, where the partial charge of
each bead was determined as the sum of the partial charges of the
underlying atoms. 
The study highlighted the interplay between
long-ranged electrostatics and the collective short-ranged
interactions between the side chain groups, along with their role in forming and
stabilizing spatially heterogeneous domains.

Bhargava \etal proposed a CG model for the family of
[C$_n$mim]$^+$ cations together with the anion hexafluorophosphate
([PF$_6$]$^-$)~\cite{Bhargava2007}. Three CG beads were devoted to the
methylimidazolium atoms to ensure the planarity of the ring, while using 
one to three CG beads to represent the alkyl chain, depending on its
length. Transferability between butyl, heptyl, and decyl side chains
was obtained by means of two bead types: terminal and interior beads. 
A single bead was used to represent the anionic PF$_6$. Both
intra- and intermolecular interactions were represented by simple
functional forms: harmonic potentials for the bonds and bending
angles; 9--6-type Lennard-Jones potentials for the short-ranged nonbonded interactions; and explicit Coulomb
electrostatics. The parameters of the model were optimized to
reproduce several properties of [C$_4$mim][PF$_6$] under ambient conditions: ($i$) the mean of reference all-atom (AA) distributions along each order parameter that governs an intramolecular interaction in the CG potential, 
($ii$) the density, and ($iii$)
the surface tension.
The authors' analysis highlighted the role of the alkyl chains in the morphology
of the liquid, leading to nanoscale ordering, in good agreement with
X-ray scattering experiments. The study remains, to date, an excellent
landmark for CG models in their capability to reproduce various
features of RTILs.

Karimi-Varzaneh \etal constructed a model that focuses on the
same chemistries, i.e., [C$_n$mim][PF$_6$] with $n = \{4,7,10\}$, and
additionally considered two mappings~\cite{KarimiVarzaneh2010}. While
the first mapping resembled the one used by Bhargava \etal,
the second one notably used a single bead to represent the imidazolium
ring. The CG model was parametrized using iterative Boltzmann
inversion, fitting all radial distribution functions while
employing a single short-ranged pairwise potential per pair type,
i.e., without employing explicit electrostatics. While only targeting
structural features, excellent agreement was also found for the
surface tension, when compared against both the CG model from Bhargava
\etal and experiments. Aspects of temperature
transferability were probed, with a focus on the change in the major
peak positions of the X-ray scattering structure factors. The choice
of mapping had a direct impact on the quality of the peak positions,
highlighting that the aromatic ring and the alkyl chains impact
different regions of the structure factor. Finally, the dynamics of
the CG models were characterized. Previous studies employing both
pulsed field gradient NMR~\cite{Tokuda2004} and AA simulations
have demonstrated that the diffusion coefficient ($D$) of the cation
is higher than that of the anion for $n=4$.
Karimi-Varzaneh \etal reported that the ratio $D_{\rm cation}/D_{\rm anion}$ depends on the mapping scheme, the value of $n$, and temperature.
Moreover, a better performance of the second mapping in terms of reproducing the non-Gaussian particle displacement statistics of the AA model highlighted the role of a more detailed representation of the alkyl chains.

More recently, Deichmann and van der Vegt revisited the bottom-up
parametrization of [C$_4$mim]$^+$ with three different anions:
[PF$_6$]$^-$, tetrafluoroborate, and chloride~\cite{Deichmann2019}.
The CG model employed a two-bead representation for the cation:
one for the imidazolium ring and one for the alkyl chain. They relied
on the conditional reversible work method, which determines the CG parameters 
from (biased) AA simulations by invoking a thermodynamic cycle.
They used Morse-type potentials for the short-ranged nonbonded interactions, 
in addition to explicit Coulomb
electrostatics. The resulting thermodynamic properties of the model---both the liquid--vapor
surface tension and mass density---agreed well with AA
simulations after an a posteriori correction of the force field.
Overall, good agreement was found in terms of the radial distribution
functions, although a few outliers were observed, in particular those involving both the
imidazolium ring and the [PF$_6$]$^-$ anion. 
Regarding dynamics, it was found that the diffusive speedup with respect to the AA model differs for the cation and anions.
We also note the development of other CG models for RTILs using
alternative methodologies: Newton Inversion~\cite{Laaksonen_PCCP_13} (also known as Inverse Monte Carlo),
relative entropy~\cite{Aluru_JCTC_18}, and graph neural networks~\cite{Ruza:2020}. 
While each of these studies features different parametrization schemes and a variety of resulting properties, they share the common difficulty of \emph{representability}---simultaneous reproduction of structural, thermodynamic, and dynamical properties.

The dynamical properties of CG RTIL models are of particular interest,
since the link between structural and dynamical heterogeneity in these
systems remains unclear. Unfortunately, the interpretation of CG
dynamics is inherently difficult, as the removal of degrees of freedom
from the system results in both a loss of friction and a ``smoothing''
of the underlying free-energy landscape~\cite{Mukherjee:2017}. Although generalized Langevin
dynamics can be applied to correct for these effects, this approach
remains extremely challenging for complex soft matter systems, and
also (partially) removes the beneficial speedup provided by the CG
model~\cite{Rudzinski2019}. Instead, many researchers have taken a much simpler approach by
determining an effective dynamical speedup factor, e.g., by
calibrating against a long-timescale dynamical observable. For
specific systems, and in particular when a clear timescale separation exists between characteristic processes,
this time-rescaling approach can be highly effective~\cite{Harmandaris:2009oc, Salerno:2016}. In general, though, the
presence of multiple coupled kinetic processes, which may be independently
accelerated by the coarse-graining, leads to inconsistent dynamics~\cite{Rudzinski:2016, Rudzinski:2016b}. A similar dynamical inconsistency typically arises 
in CG models of multicomponent systems with distinct molecular
species, e.g., cations and anions in RTILs~\cite{KarimiVarzaneh2010, Deichmann2019}. 
For bottom-up models,
improving the description of the many-body potential of mean force (i.e., the theoretically-ideal CG potential) offers a systematic
route toward one important aspect of dynamic (in particular, kinetic) consistency: barrier-crossing dynamics~\cite{Bereau:2018}. 
This link further justifies efforts to improve the structural accuracy of CG models via many-body~\cite{Larini:2010dq, Das:2012c,Lindsey:2017, Scherer:2018, John:2017, Zhang:2018, Wang:2019, Chan:2019, Scherer:2020, Ruza:2020, Shahidi:2020, Zuo:2020} or environmental-dependent~\cite{DeLyser:2017, Sanyal:2018, Jin:2018, DeLyser:2019, Shahidi:2020, Davtyan:2014, Katkar:2018, Bereau:2018, Dama:2017, Rudzinski:2020} interactions.

Recently, Pal and Vogel investigated the relationship between various dynamical modes and the relevance of spatially heterogeneous dynamics for AA and CG models of [C$_4$mim][PF$_6$]~\cite{Vogel_CPC_17}. 
They found that, despite the nontrivial transformation of dynamical processes upon coarse-graining, several relationships between dynamical modes on very different timescales are preserved for both ionic species. 
In particular, time scales of maximum non-Gaussian and spatially heterogeneous dynamical behaviors collapse onto the same curve when plotted as a function of the structural relaxation time, independent of the value of the ionic charges and, thus, of the electrostatic interactions~\cite{Vogel_JCP_19_TP}, and consistent with results for various types of viscous liquids~\cite{Vogel_JCP_15_PH}.
Somewhat related, Douglas and coworkers have put forth an energy-renormalization method for coarse-graining that makes use of a relationship between the structural relaxation time of a liquid and the temperature-dependent activation free energy, which is linked to the configurational entropy of the system~\cite{Xia:2017}.
This approach determines a temperature-dependent adjustment to the cohesive energy of the molecular interactions, which has been demonstrated to result in \emph{consistent} CG dynamics over temperature and frequency (for small-amplitude oscillatory shear molecular dynamics) for liquid ortho-terphenyl~\cite{Xia:2018} and several polymer melts~\cite{Xia:2017,Song:2018,Xia:2019}.
These studies demonstrate that by preserving certain short-timescale quantities under coarse-graining of these representative systems, the long-timescale dynamics can be predicted. 

In this study, we aim to better understand the connection between certain CG parametrization strategies and the dynamical properties and temperature transferability of the resulting models.
We systematically compare five CG models: 
\begin{enumerate}
\item ``thermo'': the model from Bhargava
\etal~\cite{Bhargava2007}, largely parametrized from experimental
thermodynamic observables; 
\item ``thermo*'': a refinement of the thermo model to increase its
structural accuracy; 
\item ``struct-anabond'': a structure-based model focused on
reproducing intermolecular distributions; 
\item ``struct'': a structure-based model that targets both intra-
and intermolecular distributions; 
\item ``struct-260K'': a structure-based model that targets both
intra- and intermolecular distributions, parametrized at a lower
reference temperature than the struct model.
\end{enumerate}
All five CG models display limited structural transferability over temperature, and also result in various effective dynamical speedup factors relative to a reference AA model.
For each model, we quantify the speedup, analyze the relative diffusivity of cations and anions, and study the relationship between vibrational and diffusive motions. 
In this way, we provide further evidence of a route to constructing dynamically-consistent CG simulation models, in particular for RTILs.

\section{Methods}

\subsection{All-atom simulations}

The present work makes use of AA simulation trajectories of [C$_4$mim][PF$_6$] generated and analyzed in previous works~\cite{Vogel_JPCC_18, Vogel_CPC_17, Vogel_JCP_19_TP}. 
Briefly, the force field from Bhargava and Balasubramanian~\cite{Bhargava2007b} was used, which features cations and anions with partial charges of $+0.8e$ and $-0.8e$, respectively, harmonic potentials for bond stretching and bond-angle bending as well as dihedral interactions. The simulations were performed using the GROMACS software package~\cite{GROMACS5} for $N=256$ ion pairs and a time step of 1\,fs. Periodic boundary conditions were applied and the Particle Mesh Ewald (PME) method~\cite{Darden:1993} was utilized to calculate the Coulomb interactions. At all set temperatures $T$, the system was first equilibrated at a pressure $P=1$\,bar to adjust the density, employing the Nos\'e-Hoover thermostat~\cite{NH1, NH2} and Parrinello-Rahman barostat~\cite{Parrinello:1982}. The production runs were carried out in the isochoric-isothermal ($NVT$) ensemble.
We note that, as can be seen in Fig.~\ref{fig:msd}, the simulations at the lowest temperatures (i.e., below 300~K) may not be fully equilibrated, as the particles do not completely reach the diffusive regime.
For this reason, we do not include the data from these temperatures in the dynamical property analysis.
Nonetheless, we have used the lowest temperature simulation to construct one of the CG models (see below), based on the structural properties, which presumably are much less sensitive to the full equilibration of these simulations.
Additionally, we use the structural properties from these lower temperatures to compare to the properties of the CG models.

\subsection{Coarse-grained representations and interactions}

We employ the CG mapping proposed by Bhargava \etal~\cite{Bhargava2007} and used previously~\cite{Vogel_CPC_17}, which represents each imidazolium cation with 4 CG sites and each [PF$_6$]$^-$ anion with a single CG site.
The imidazolium ring is represented by 3 sites, I1, I2 and I3, mapped to the center of mass of a corresponding group of atoms, as illustrated with the large transparent spheres in Fig.~1.
Note that the I1 and I2 sites overlap, sharing contributions from the 2-carbon of the 5-membered ring (i.e., the carbon flanked by two nitrogens).
The butyl chain is represented by an additional site, denoted CT, while the anion site is denoted PF.
In the following, this mapping is applied to 5 different CG models, each employing the same set of interactions, e.g., nonbonded pairwise interactions between all unique pair types, but with variations in the functional forms and parameters of these interactions.
4 bonded interactions are employed to retain the imidazolium connectivity: I1--I2, I1--I3, I2--I3, I2--CT.
Accordingly, each model employs 5 bond-angle interactions: I1--I2--I3, I1--I3--I2, I3--I1--I2, I1--I2--CT, I3--I2--CT.
There are no dihedral angle interactions in these models.
Finally, 15 pairwise nonbonded interactions, corresponding to all unique pair combinations of the 5 site types, are employed, while excluding intramolecular imidazolium pairs. 
There are two distinct contributions to each nonbonded interaction: (i) a short-ranged van der Waals-like interaction that is parametrized separately for each model, and (ii) a long-ranged Coulomb interaction that is kept fixed for all models.
The Coulomb forces are calculated by mapping the partial charges of the AA model to the CG representation: $q = $ $+0.356e$, $+0.292e$, $+0.152e$, $0e$, and $-0.8e$ for the I1, I2, I3, CT and PF sites, respectively. 

\begin{figure}[htbp]
        \begin{center}
                \includegraphics[width=0.5\linewidth]{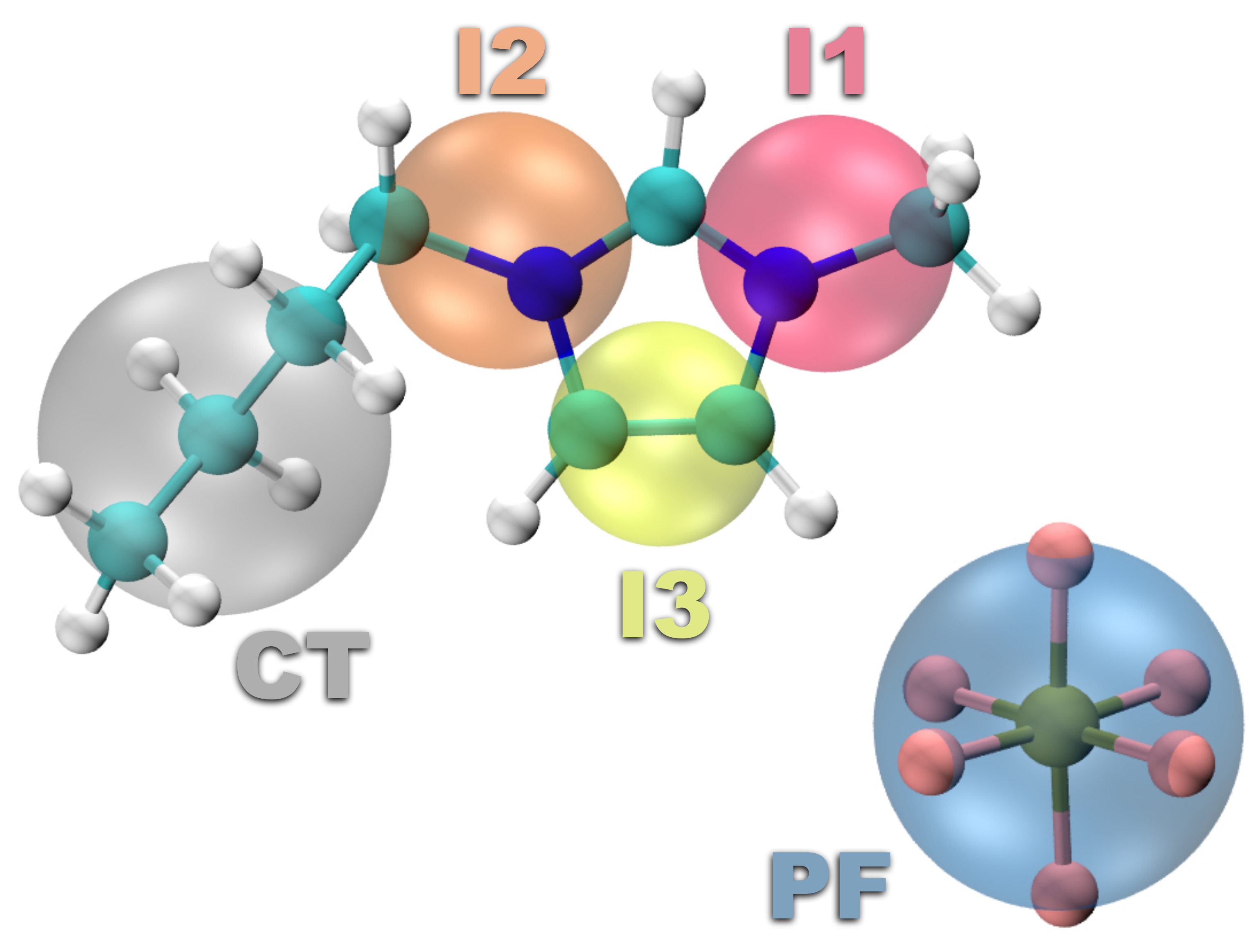}
                \caption{
CG representation.
Each molecule is partitioned into a number of atomic groups, as indicated by the large transparent spheres: 4 for the cation and 1 for the anion.
All CG beads were mapped to the center of mass of the corresponding group of atoms:
($i$) CT -- the last 3 carbon groups of the alkyl chain;
($ii$) I2 -- the first carbon group of the alkyl chain, the 1-nitrogen of the ring, and ``half'' of the 2-carbon group of the ring;
($iii$) I1 -- the 3-nitrogen and associated methyl group of the ring and also ``half'' of the 2-carbon group of the ring;
($iv$) I3 -- the 4- and 5-carbon groups of the ring; 
($v$) PF -- the entire anion.
}
                \label{fig:CG-rep}
        \end{center}
\end{figure}

\subsection{Coarse-grained model parametrizations}

\subsubsection{Methods}

\hfill\\

\noindent \uline{Iterative Boltzmann Inversion (IBI):}
IBI is a bottom-up method that determines the CG potential that will reproduce a given set of 1-D distribution functions through an iterative refinement that assumes independence of the CG potential parameters~\cite{Soper:1996ly,Reith:2001}.
Given an initial model and a set of reference distributions, each interaction potential, $U$, is updated at each iteration according to:

\begin{equation}
        \label{eq-IBI}
        U^{(i+1)}(\xi) = U^{(i)}(\xi) - \alpha \kT \ln \frac{P^{(i)}(\xi)}{P^{(\rm ref)}(\xi)}      \,\, ,
\end{equation}
where $k_B$ is the Boltzmann constant, $\xi$ is a scalar order parameter that governs the interaction, and $P(\xi)$ is the Jacobian-transformed distribution function along $\xi$.
For pairwise nonbonded interactions, $P(\xi)$ corresponds to the radial distribution function (RDF), $g(r)$, where $r$ is the distance between two particles.
$\alpha \in [0,1]$ is a damping factor used to increase the stability of the procedure. 

\noindent \uline{Iterative generalized Yvon-Born-Green (iter-gYBG)}:
We also considered an alternative iterative parametrization scheme, 
using a generalization of the Yvon-Born-Green (YBG) integral equation framework~\cite{Mullinax:2009b,Mullinax:2010}.
This approach relates a set of structural correlation functions, $\bb$, to the parameters of the CG interaction potentials, $\bphi$, via a matrix equation, which quantifies the cross-correlations, $\bG$, between each CG degree of freedom:

\begin{equation}
        \label{eq-gYBG}
        \bb = \bG \bphi      .
\end{equation}
For nonbonded, pairwise interactions represented by a set of spline functions, $\bb$ is directly related to the corresponding RDF, and $\bG$ characterizes the average cosine of the angle between triplets of particles~\cite{Rudzinski:2012vn,Rudzinski:2015}.
$\bb$ can also be expressed in terms of force correlation functions through a connection to the multiscale coarse-graining method~\cite{Mullinax:2010, Noid:2008a}.
This method employs force and structural correlation functions that are determined from a set of AA reference simulations, $\bb^{\rm AA}$ and $\bG^{\rm AA}$, i.e., calculated from a set of AA simulation trajectories mapped to the CG representation, to determine an optimal set of CG parameters $\bphi$.
Note that if the model derived from this method fails to reproduce the target vector of
the equations, i.e., $\bb^{\rm AA}$, it implies that the cross-correlation matrix generated by the higher resolution model does
not accurately represent the correlations that would be generated by
the resulting CG model.  This indicates a fundamental limitation of the model representation and interaction set.
Nonetheless, 
Equation~\ref{eq-gYBG} can be solved self-consistently to determine the interaction parameters $\bphi^*$ that reproduce the target correlations $\bb^{\rm AA}$~\cite{Rudzinski:2014}:
\begin{equation}
        \label{eq-itergYBG}
        \bb^{\rm AA} = \bG(\bphi^*) \bphi^* .
\end{equation}
This approach has been previously denoted as an iterative generalized
YBG (iter-gYBG) method~\cite{Rudzinski:2014, Cho:2009ve,Lu:2013uq}.
When the multiscale coarse-graining model is employed as the initial
set of parameters, this procedure has been demonstrated to converge
very quickly (e.g., in less than 10 iterations), although it may be
less robust than IBI~\cite{Rudzinski:2014,Rudzinski:2014b}.

\subsubsection{Models}

\hfill \\

\noindent \uline{thermo:}
As already mentioned in the Introduction, the model proposed by Bhargava
\etal~\cite{Bhargava2007} was parametrized in a top-down fashion to
reproduce experimental thermodynamic properties, and will thus be
denoted as "thermo" in the following. This model employs analytic
functional forms for the CG interaction potentials: harmonic
potentials for intramolecular interactions and 9--6 Lennard-Jones 
potentials for the nonbonded interactions. Initial equilibrium
distances and force constants for the intramolecular potentials were
determined by fitting the Boltzmann-inverted potentials from AA
simulations to the harmonic functional form.
These parameters were then refined to reproduce the mean of the 1-D
distributions along each order parameter governing an intramolecular interaction in the CG model.
The Lennard-Jones energies and particle radii were then tuned
to reproduce the density and surface tension from experimental
measurements at 300~K. It is also reported that the site--site RDFs
from AA simulations were taken into account when determining these
parameters, although it is not clear to what extent or how this was
carried out. 

\noindent \uline{thermo*:}
We performed a refinement of the thermo model to improve its structural accuracy, while attempting to minimally perturb the thermodynamic properties of the model.
To achieve this, we applied the iter-gYBG approach to each of the short-ranged nonbonded interactions, but restricted our calculation to relatively short distances between CG sites.
More specifically, we calculated the change in the force coefficients up to the distance corresponding to the minimum of each of the potentials (see Supporting Information and manuscript repository~\cite{Rudzinski:2021-repo} for details).
We also calculated the change in the intramolecular interactions, but ignored these when updating the force field at each step.
Previous investigations have shown that taking into account the coupling between intra- and intermolecular interactions can have a substantial impact on the resulting force functions from the g-YBG approach (unpublished). 
We also performed a linear interpolation between the old and new forces over a range of 0.05~nm preceding the cutoff distance used for the force field calculation, followed by a smoothing of the resulting force functions to avoid unphysical kinks.
The procedure was terminated after 6 iterations, since further iterations resulted in numerical artifacts, perhaps due to the strong restriction of the tunable force field parameters.
Further details of the iter-gYBG calculations can be found in the description of the parametrization of the struct model below.

\noindent \uline{struct-anabond:}
The struct-anabond model was parametrized using IBI to adjust the short-ranged nonbonded interactions to reproduce the RDFs from the reference AA simulation at 300~K, while keeping the intramolecular interactions and long-ranged electrostatics fixed to those used in the thermo model.
The procedure was terminated after 160 iterations (see manuscript repository~\cite{Rudzinski:2021-repo} for details).
IBI calculations were performed using the VOTCA package~\cite{Ruhle:2009wx}.

\noindent \uline{struct:}
The struct model was parametrized with the iter-gYBG method, using the reference AA simulations at 300~K and the BOCS package~\cite{Dunn:2018}.
All bond and bending-angle force functions were represented with linear spline basis functions with grid spacings of 0.001~nm and 1~deg, respectively.
The short-ranged nonbonded force functions were represented with B-spline basis functions with a grid spacing of 0.01~nm.
The initial model was determined by solving Eq.~\ref{eq-itergYBG} while using the cross correlations from the AA simulation to compute $\bG$ (i.e., the initial model was obtained via the multiscale coarse-graining method).
The iter-gYBG model was then calculated following the iter-gYBG framework, by solving Eq.~\ref{eq-itergYBG} iteratively, using cross correlations determined from the CG simulations at the previous step, until a pre-set accuracy threshold was achieved with respect to the 1-D distribution functions along each order parameter governing a CG interaction (8 iterations in this case).
The first 3 iterations scaled the calculated parameter update by a factor of 0.25, while further iterations applied a factor of 0.5.
For these calculations, the \emph{fixed} long-ranged Coulomb interactions were incorporated via the reference potential method~\cite{Mullinax:2010a}, which subtracts the contributions to the force correlations, $\bb$, from a given set of fixed terms in the force field.
To increase numerical stability, we also used the direct Boltzmann inverted forces for each intramolecular interaction as reference, even though we still calculate the optimal forces for these interactions (i.e., we calculate the change in the force parameters, relative to the reference forces).
Solution of each set of linear equations and post-processing of the potentials followed previous work~\cite{Rudzinski:2014, Rudzinski:2014b}.

\noindent \uline{struct-260K:}
The struct-260K model was parametrized in the same way as the struct model, but using the reference AA simulations at 260~K.
In this case, 7 iterations were required to achieve the accuracy threshold.

\subsection{Coarse-grained simulations}

All CG simulations were performed using the GROMACS simulation
package~\cite{Hess:2008}. Each simulation, with the exception of those associated with the thermo model (see further details below), was performed at the same density as the corresponding AA simulation of the same temperature (see Table~S2 for the specific densities and comparisons to experiments).
Starting from an equilibrated AA configuration mapped to the CG
representation, each CG model was applied to energy minimize the
configuration, followed by a 10~ns simulation in the $NVT$ ensemble at the respective temperature.
All structural analysis was performed using these simulations.
For the dynamical analysis, some models/temperatures required longer simulations.
In particular, for the thermo* model, we performed additional 25~ns simulations at 400, 350, 320, and 300~K and additional 100~ns simulations at 280, 270, and 260~K.
Similarly, for the struct-anabond, struct and struct-260K models, we performed additional 25~ns simulations at 280, 270, and 260~K.
All CG simulations used the stochastic dynamics integrator
with a temperature coupling constant of 2~ps, a 1~fs time step, and
periodic boundary conditions. Electrostatic interactions were employed
using the PME method~\cite{Darden:1993} with a Fourier grid spacing of
0.10~nm. A cutoff of 1.5~nm was used for both the short-range
nonbonded interactions and for the real-space contribution to
electrostatic interactions. Configurations were sampled every 0.1~ps
during the simulations, and then used for subsequent dynamical
analysis. The structural analysis was performed using trajectories
parsed to yield configurations every 0.4~ps.

The dynamical properties of the thermo model were taken from previously published simulations~\cite{Vogel_CPC_17}.
These simulations were carried out at slightly different densities than the AA model (see Table~S2), corresponding to the equilibrium density of the model at 1~bar.
To assess the impact of this change in density on the model properties, we ran 10~ns simulations of the thermo model at each temperature and corresponding AA density, following the protocol described above.
All structural analysis presented in this work was taken from these simulations for consistent comparisons with the other models.
We note that the slight change in density has negligible impact on the RDFs for nearly all of the simulations.
At the lowest temperature (260~K) there are noticeable, albeit very small, deviations in a few of the RDFs.
(The full set of RDFs are available in the manuscript repository~\cite{Rudzinski:2021-repo} for comparison).
Similarly, we expect only a relatively small impact on the dynamics, although we did not explicitly verify this.

Throughout this manuscript we will describe the dynamical timescales
of each CG model in terms of physical time units, in order to achieve
a consistent comparison with the AA model. However, as described in
the Introduction, the process of coarse-graining results in a lost
connection between the CG and AA dynamics. Thus, the reported CG
timescales are not meaningful without a proper reference or
calibration. However, comparison of relative timescales can assist
in assessing if a CG model exhibits consistent dynamics compared with
the AA model~\cite{Rudzinski:2016}.

\section{Results}

\subsection{CG Models}

In this work, we investigate a spectrum of particle-based CG models for [C$_4$mim][PF$_6$], each parametrized to target specific structural or thermodynamic observables of the underlying system, characterized either with respect to experiments or using AA simulations.
We consider five models---denoted thermo, thermo*, struct-anabond, struct and struct-260K---as described in detail in the Methods section.
In brief, the thermo model was previously parametrized to reproduce the experimental density and surface tension~\cite{Bhargava2007}.
The thermo* model is a refinement of the thermo model, which more accurately reproduces the anion-cation radial distribution functions (RDFs) by adjusting the (very) short-ranged region of the nonbonded interactions, while keeping the remaining interactions fixed.
By focusing on adjusting the force functions at short distances, the thermo* model largely retains the thermodynamic accuracy of the original (thermo model) parametrization.
In particular, the equilibrium density of the thermo* model at 300~K was determined to be 1.301~g/cm$^3$, compared with 1.357~g/cm$^3$ for the thermo model, 1.389~g/cm$^3$ for the AA model, and 1.369~g/cm$^3$ from experiments.
Additionally, the surface tension of the thermo* model at 300~K was determined to be 38.1~mN/m, compared with 39.0~mN/m for the thermo model and 42.5~mN/m from experiments (see Supporting Information for calculation details).
(Note that, for consistent comparisons in the following analysis, the CG models were simulated at the corresponding AA equilibrium density. However, the dynamical analysis of the thermo model, taken from previously published work~\cite{Vogel_CPC_17}, employed slightly different densities. See the Methods section for details.)
The struct-anabond model uses the analytic (harmonic) functional forms and parameters for the intramolecular interactions from the thermo model, while employing tabulated potentials for the short-ranged nonbonded interactions to reproduce the RDFs from reference AA simulations at 300~K.
The struct model was also parametrized to reproduce these same RDFs, but additionally employs tabulated intramolecular interactions to reproduce the corresponding 1-D distributions along each CG (intra)molecular degree of freedom.
Finally, the struct-260K model is equivalent to the struct model, but parametrized using the reference distributions at 260~K.

\begin{figure}[htbp]
        \begin{center}
                \includegraphics[width=\linewidth]{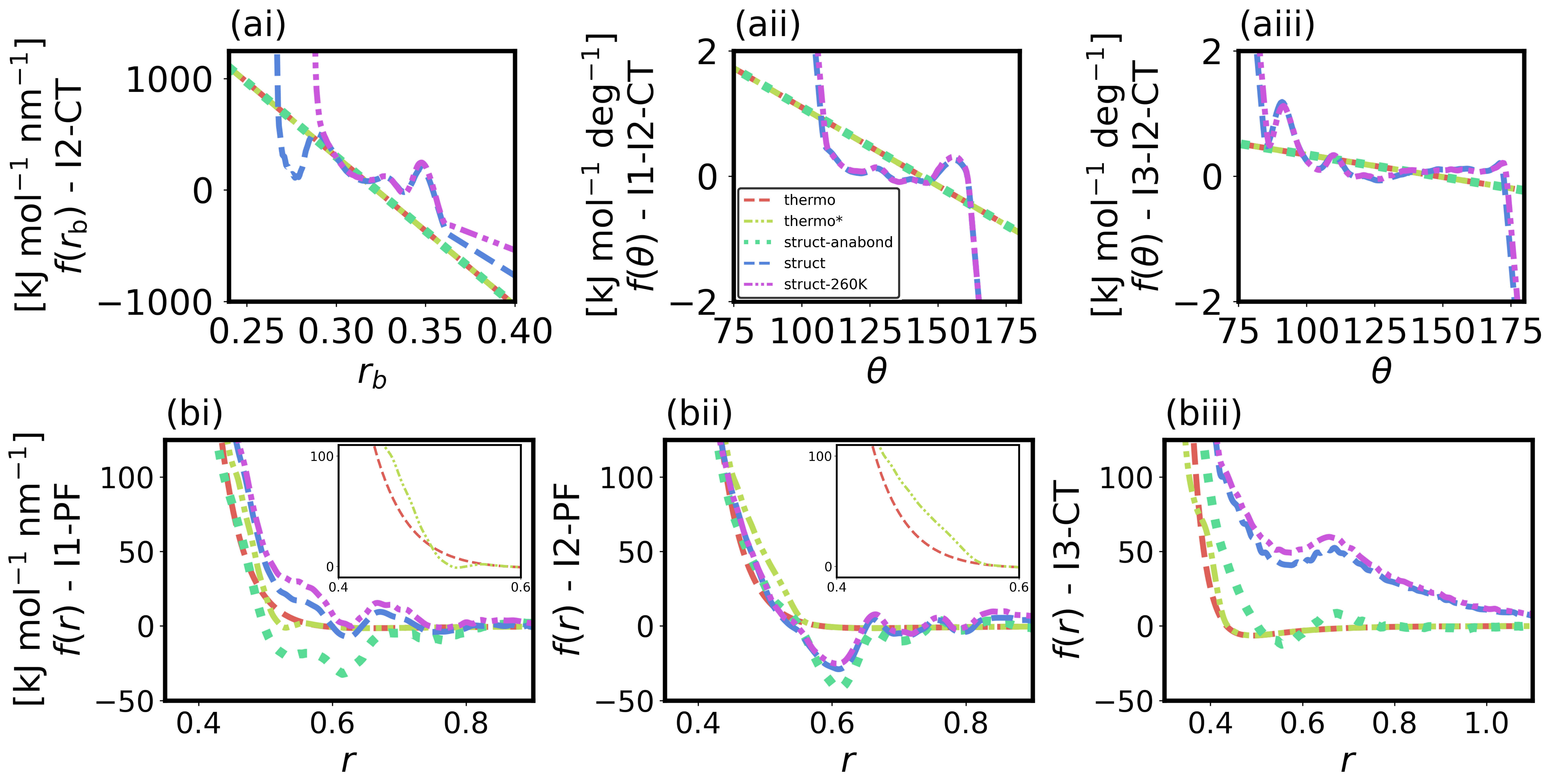}
                \caption{
Comparison of representative intra- (a) and inter- (b) molecular force functions for the thermo (red dashed curve), thermo* (yellow dash-dotted curve), struct-anabond (green dotted curve), struct (blue dashed curve), and struct-260K (purple dash-dotted curve) models.
The full set of force functions is presented in Figs.~S1--S3 of the SI. 
}
                \label{fig:cf-U}
        \end{center}
\end{figure}

Fig.~\ref{fig:cf-U} compares the force functions for each model, for a representative set of interactions.
Row (a) presents the only three intramolecular interactions whose corresponding 1-D reference distributions (i.e., calculated from the AA simulations after mapping each configuration to the CG representation) displayed multimodal behavior.
It is worth noting that the I1--I2--CT and I3--I2--CT interactions are significantly coupled to one another.
As mentioned above, the thermo, thermo* and struct-anabond models (red, yellow and green curves, respectively) employ identical intramolecular force functions, in this case a linear function corresponding to the harmonic interaction potential form.
The struct and struct-260K models (blue and purple curves, respectively), which employ tabulated force functions to match the corresponding 1-D distribution functions, demonstrate overall weaker forces, within the central interaction regime, than their analytic counterparts.
The linear extrapolation of these forces to large values at the ends of the interaction range is largely arbitrary and does not impact the accuracy of the resulting distribution (within some reasonable range).

Row (b) presents the three short-range nonbonded interactions corresponding to the RDFs with the largest errors in the thermo model (described further in the following section).
The thermo model (red curve) employs a 9--6 Lennard-Jones functional form with potential minima of $\approx$~0.35~kJ/mol for the I1--PF and I2--PF interactions and $\approx$~1.2~kJ/mol for the I3--CT interaction.
The thermo* model (yellow curve) was constructed by adjusting the hard core of each short-range nonbonded interaction (specifically, for distances shorter than the location of the potential minimum), in an attempt to increase the structural accuracy of the model while minimally perturbing the thermodynamic properties.
For example, the I1--PF and I2--PF interactions were refined for distances shorter than 0.584~nm while the I3--CT interaction was refined for distances shorter than 0.435~nm.
These adjustments had the largest impact on the I1--PF and I2--PF force functions, as can be seen more clearly in the insets of panels (bi) and (bii).
The struct-anabond, struct, and struct-260K models (green, blue and purple curves, respectively) each employ tabulated force functions to reproduce a given set of RDFs.
The resulting forces tend to have significantly larger magnitude than the forces of the thermo model, along with more complicated features, as expected.
The struct and struct-260K models tend to be much more similar than the struct-anabond model, as is clear from Fig.~\ref{fig:cf-U}(b).
This is almost certainly because the former two models are both determined from just a few iterations (i.e., less than 10) starting from the force-matching-based multiscale coarse-graining model at each temperature, while the struct-anabond model is determined from over 100 iterations starting from the thermo model (see Methods section for details).
All three of these models demonstrate large pressures in the $NVT$ simulations that prohibit the characterization of the typical thermodynamics properties, e.g., density and surface tension, as is common for structure-based CG models of liquids~\cite{Johnson2007,Wang:2009ol,Guenza:2015,Dunn:2016b}.

\subsection{Structural Properties}

\begin{figure*}[htbp]
        \begin{center}
                \includegraphics[width=\linewidth]{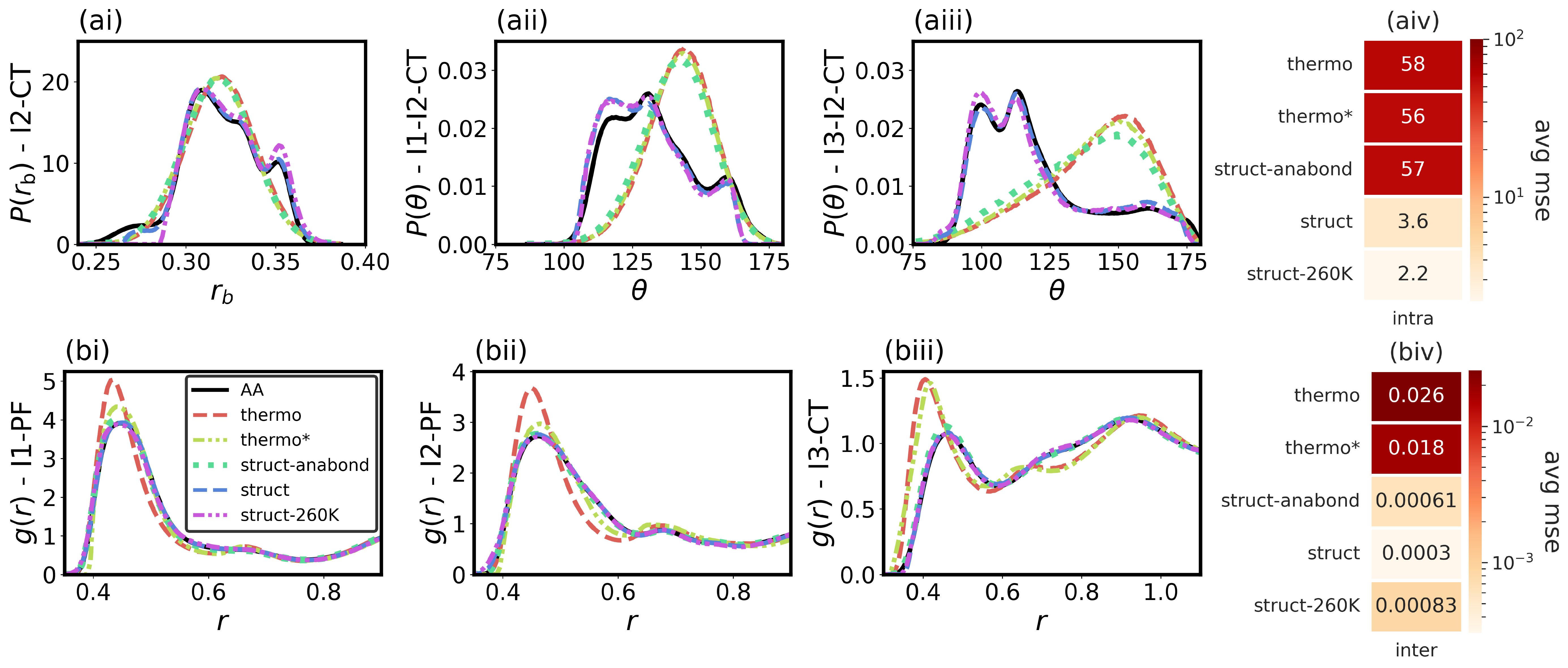}
                \caption{
Comparison of representative intra- (a) and inter- (b) molecular 1-D distributions at 300~K for the AA (black solid curve), thermo (red dashed curve), thermo* (yellow dash-dotted curve), struct-anabond (green dotted curve), struct (blue dashed curve), and struct-260K (purple dash-dotted curve) models.
The full set of distributions is presented in Figs.~S4-S6 of the SI.
                }
                \label{fig:struct-300}
        \end{center}
\end{figure*}

Figure~\ref{fig:struct-300} presents the 1-D distributions
corresponding to the interactions presented in Fig.~\ref{fig:cf-U}.
Panel (ai) shows that the harmonic potential of the thermo model leads
to an accurate description of even the most complicated bond
distribution in this molecule, the I2-CT bond, although the fine
features (e.g., shoulders) of the distribution are not reproduced. The
struct model nearly quantitatively reproduces the distribution, by
construction. The small remaining discrepancies are due to numerical
difficulties of the iter-gYBG procedure which can occur at the tails
of the distribution. This issue could likely be resolved by employing
a regularization scheme, but we do not
pursue this here, as we don't expect these discrepancies to affect our
results. The struct-260K model matches this distribution for most of
its range, but completely neglects the small shoulder at short
distances, due to this same numerical issue. Panels (aii) and (aiii)
demonstrate that the angular distributions involving the alkyl chain
of the cation are quite complex, and cannot be accurately described
with the harmonic potential employed by the thermo model. The struct
and struct-260K models reproduce these distributions by construction,
albeit with some small discrepancies at the tails of the
distributions. Panel (aiv) presents the average mean squared error
(mse) for each model over all intramolecular distributions along the
CG degrees of freedom that govern an interaction in the CG model (see
Fig.~S4 for explicit comparisons of each distribution).
For each intramolecular distribution, the mse was calculated over a range encompassing all sampled instances of the corresponding order parameter, e.g., bond distance. 

Row (b) presents three sets of RDFs, demonstrating the improvement of the thermo* model, relative to the thermo model, in terms of the description of the anion-cation packing.
The remainder of the RDFs show similar behavior between the two models, analogous to panel (biii).
The struct-anabond, struct, and struct-260K models reproduce all RDFs by construction.
Panel (biv) presents the average mse for each model over all 15 RDFs (see Figs.~S5 and S6 for explicit comparisons of each RDF).
For each RDF, the mse was calculated from 0 to 1.2~nm.

\begin{figure}[htbp]
        \begin{center}
                \includegraphics[width=0.65\linewidth]{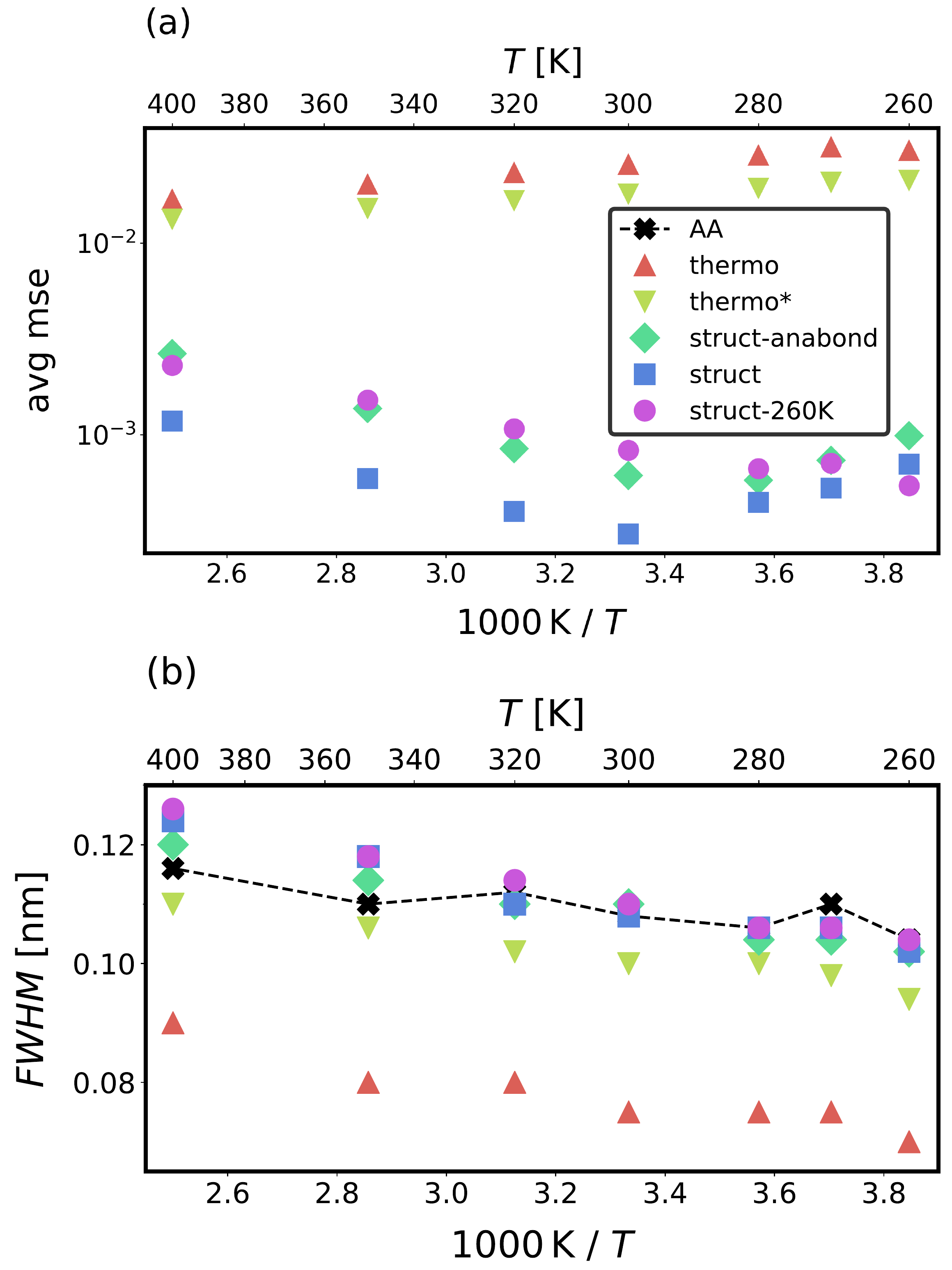}
                \caption{
Structural accuracy as a function of temperature.
(a) Average mean-squared error (mse) of the RDFs relative to the AA model.
(b) Full width at half maximum (FWHM) of the first peak of the I1-PF RDF for each model.
}
                \label{fig:T-dep-struct}
        \end{center}
\end{figure}

We also analyzed the structural properties of the models as a function
of temperature. Fig.~\ref{fig:T-dep-struct} characterizes the
structural accuracy of each model, with respect to the reference AA
model, in terms of the average mse of the RDFs (panel a) and the full
width at half maximum (FWHM) of the first peak in the I1-PF RDF (panel
b). 
Although it is somewhat difficult to see on the logarithmic scale, 
panel (a) demonstrates a decrease in structural accuracy for the 
thermo and thermo* models upon cooling, despite these models being parametrized from 
reference data at 300~K.
On the other hand, the
structure-based models (struct-anabond, struct, struct-260K), with
much lower errors overall, show divergence of the error away from the
reference temperature of parametrization (300~K for struct-anabond and
struct; 260~K for struct-260K). 
Panel (b) clearly demonstrates an improved accuracy of the thermo*
model's description of the I1--PF RDF over the entire temperature
range, relative to the original thermo model. The struct-anabond and
struct models very accurately reproduce the AA FWHM for temperatures
near the reference temperature of parametrization, but demonstrate
increasing error for the much higher temperatures. The struct-260K
model provides a similarly accurate description of the I1--PF FWHM,
with slightly larger errors at the highest temperatures, further
indicating a limited range of temperature transferability for these
structure-based models.

\subsection{Dynamical Properties}

To investigate the molecular dynamics of the AA and CG models, we calculate the mean-square displacement (MSD) of the cations and anions based on the time-dependent bead coordinates ${\bm r}_i(t)$,
\begin{equation}
    \langle r^2(t)\rangle=\left\langle\frac 1 N \sum_i^N[{\bm r}_i(t+t_0)-{\bm r}_i(t_0)]^2\right\rangle,
\end{equation}
where the pointed brackets denote an average over various time origins $t_0$. From the long-time diffusive regime of the MSD, we determine the self-diffusion coefficients $D$ of the cations and anions by fits to $\langle r^2(t)\rangle=6Dt$.
We note that the AA model generates diffusion constants in good agreement with experiments (see Table~S3 in the Supporting Information).

\begin{figure}[htbp]
        \begin{center}
                \includegraphics[width=0.6\linewidth]{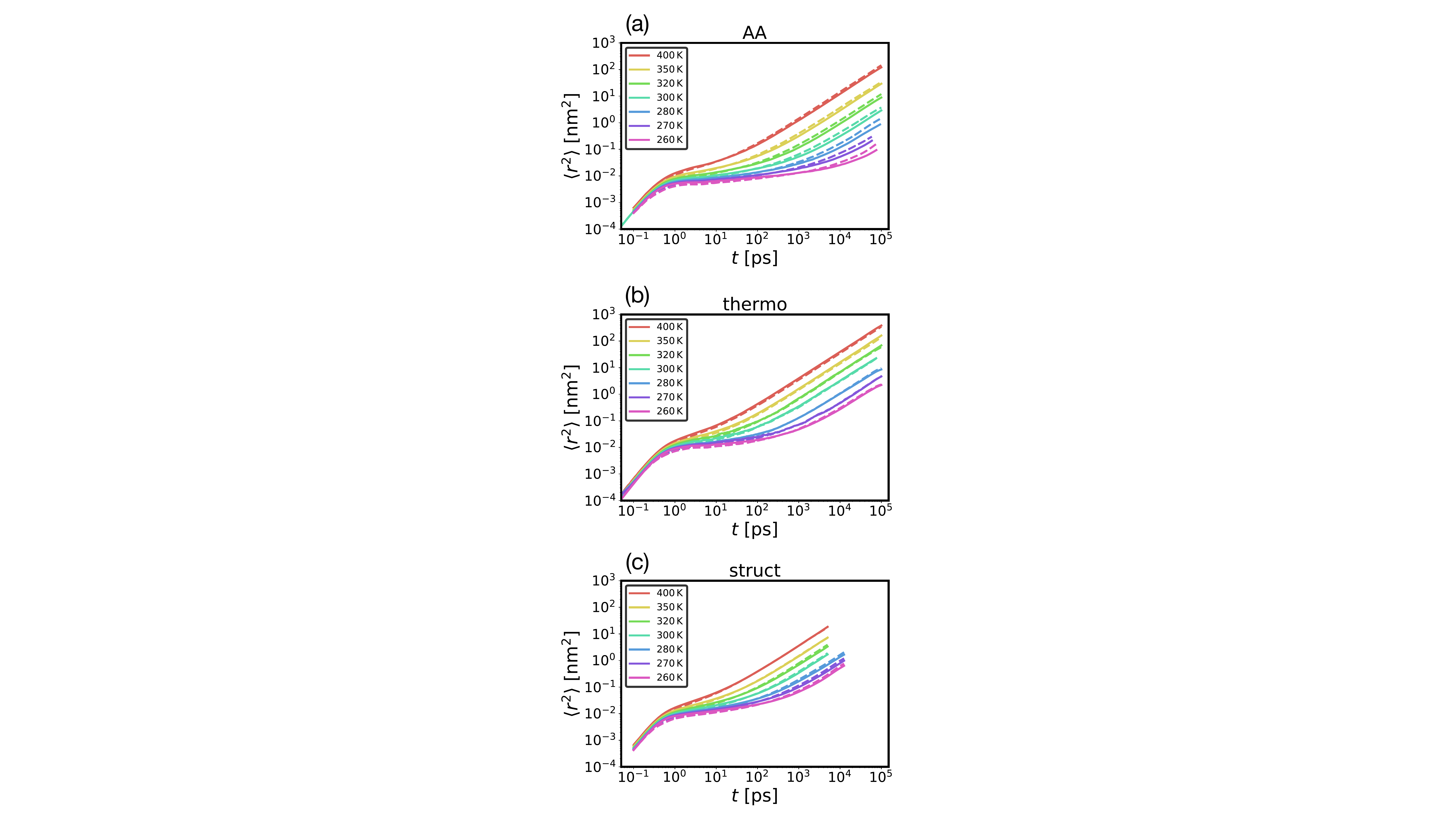}
                \caption{
Mean squared displacement (MSD), $\langle r^2 \rangle$, as a function of time for the (a) AA, (b) thermo, and (c) struct models. Results for the cations and anions are shown as dashed lines and solid lines, respectively.
}
                \label{fig:msd}
        \end{center}
\end{figure}

Figure~\ref{fig:msd} presents the MSD of the AA model and two representative CG models (the thermo and struct models) at various temperatures. 
Typical of viscous liquids, a regime of vibrational motion at short times and a regime of diffusive motion at long times are separated by a plateau, which becomes longer upon cooling~\cite{Robbins:1988}. 
This plateau regime occurs when the ions are temporarily trapped in cages formed by their neighbors and, as a consequence, display only local rattling motions. 
While the MSD is qualitatively similar for the three models, we observe quantitative differences. 
In particular, the diffusive motion has a higher temperature dependence in the AA model than in the CG models.
This is consistent with previous work~\cite{Xia:2017}, and is due to the inherent difference in energy-enthalpy compensation in the CG models~\cite{Dunn:2016b}.
Panel (a) demonstrates for the AA model that the cations (dashed curves) show higher displacements than the anions (solid curves) in the diffusive regime, consistent with experimental results~\cite{Tokuda2004}.
On the other hand, the thermo model (panel (b)) exhibits relatively similar diffusivity of the ions, with slightly larger cation displacement at low temperatures but a switch-over to lower cation displacement at high temperatures.
The thermo* model demonstrates qualitatively similar behavior (see Fig.~S25).
The struct model (panel (c)) provides a slightly more consistent description of the relative ion diffusivity, albeit with an underestimate of the cation versus anion displacements.
The struct-anabond and struct-260K models demonstrate qualitatively similar behavior (see Fig.~S25).
We note that the diffusive regime is not fully reached for the AA model at the lower temperatures of the studied range, interfering with a determination of reliable diffusion coefficients $D$. 
Therefore, we restrict the following quantitative analysis to temperatures $T\geq300$\,K.    

\begin{figure}[htbp]
        \begin{center}
                \includegraphics[width=0.6\linewidth]{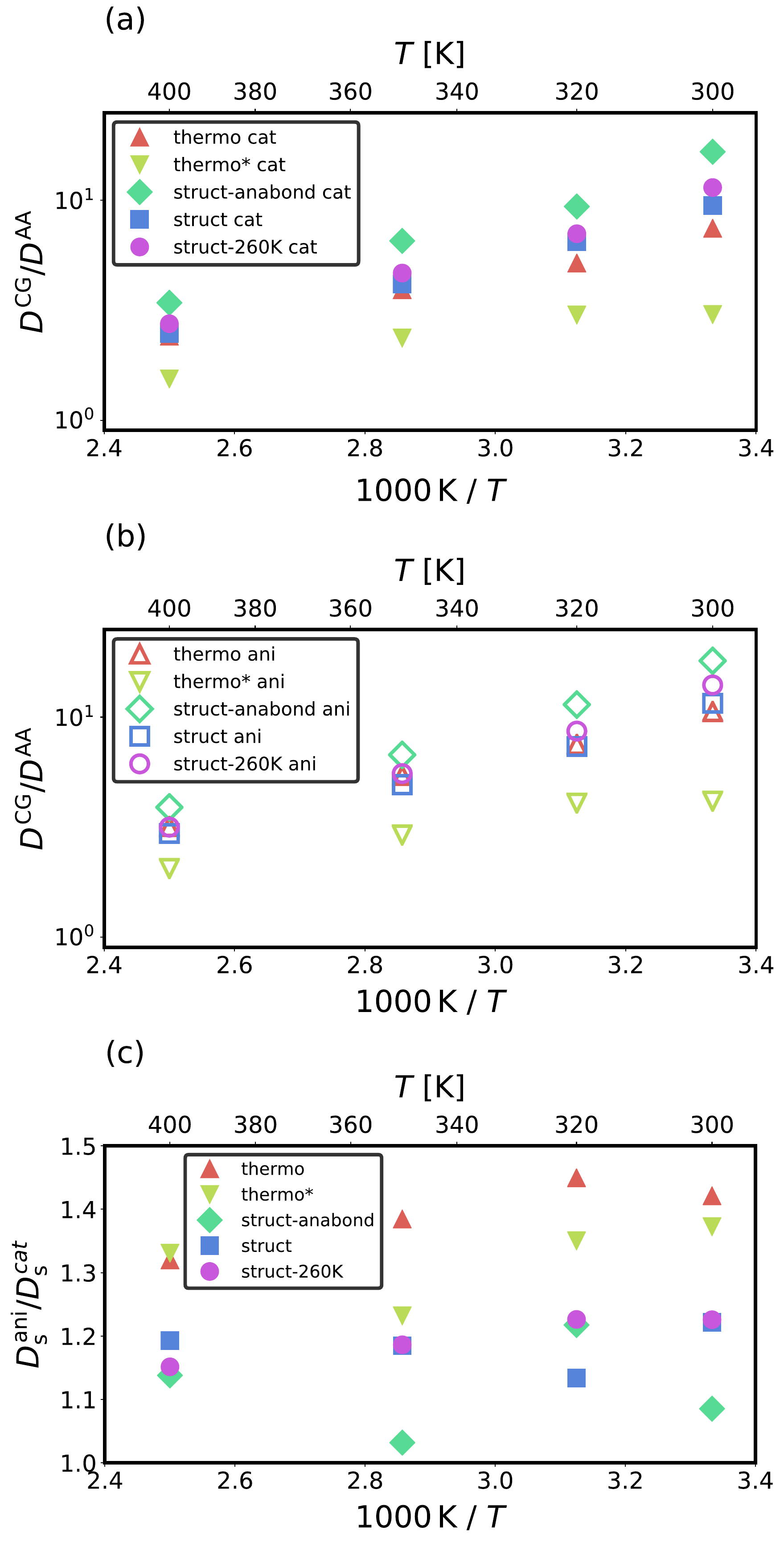}
                \caption{
Speed-up factor of the diffusion coefficient, $D_{\rm s} = D^{\rm CG} / D^{\rm AA}$, for (a) cations and (b) anions.
(c) Relative speedup of anions versus cations, $D^{\rm ani}_{\rm s} / D^{\rm cat}_{\rm s}$, at a given thermodynamic state point.
}
                \label{fig:D-speedup}
        \end{center}
\end{figure}

To quantitively compare the diffusive behavior of the different models, we calculated the diffusion coefficients for the anions and cations, using the MSD curves presented in Fig.~\ref{fig:msd}.
Fig.~\ref{fig:D-speedup} presents the temperature-dependent ratio of these coefficients for each CG model, relative to the AA model: $D_{\rm s}=D^{\rm CG}/D^{\rm AA}$.
These ratios, presented for the cations and anions separately in panels (a) and (b), respectively, represent an effective speedup factor for connecting the CG and AA dynamics, at a particular thermodynamic state point and for a specific molecular species.
As one may expect based on the smoothening of the energy landscapes upon coarse-graining, the speedup factors $D_{\rm s}$ are significantly larger than unity.
All CG models also demonstrate an increase in $D_{\rm s}$ upon cooling, consistent with the above observation of the difference in temperature dependence of the MSDs between the AA and CG models. 
By refining the structural accuracy of the thermo model, the thermo* model results in slower dynamics overall, and displays the smallest speedup factors of all the models.
Similarly, by refining the intramolecular structure relative to the struct-anabond model (which displays the largest $D_{s}$ values), the struct model also exhibits slower dynamics.
On the other hand, the change in the reference state of parametrization between the struct and struct-260K models appears to have minimal impact on the effective dynamical speedup.
Although the thermo* model yields the best \emph{absolute} reproduction of the AA dynamics, according to this metric, the more relevant quantity for assessing the dynamical consistency of each CG model is the \emph{relative} mobility of the distinct ionic species.
Panel (c) of Fig.~\ref{fig:D-speedup} presents the ratio of speedup factors between the anions and cations, $D_{\rm s}^{\rm ani} / D_{\rm s}^{\rm cat}$, at each thermodynamic state point.
In this plot, a value of unity indicates consistent CG dynamics at a single temperature, since the anions and cations experience the same speedup upon coarse-graining.
The structure-based models (struct-anabond, struct, and struct-260K) demonstrate slightly more consistent dynamics, i.e., lower $D_{\rm s}^{\rm ani} / D_{\rm s}^{\rm cat}$, over the entire range of temperatures.
Interestingly, the struct-anabond model appears to show the most consistent dynamics, despite having the largest absolute speedup factors, although the scattering of the speedup ratio is slightly larger than for the other structure-based models.
The thermo and thermo* models display larger speedup ratios, which increase slightly upon cooling.
By refining the structural accuracy relative to the thermo model, the thermo* model does result in slightly more consistent dynamics for all but the highest temperature.


\begin{figure}[htbp]
        \begin{center}
                \includegraphics[width=0.65\linewidth]{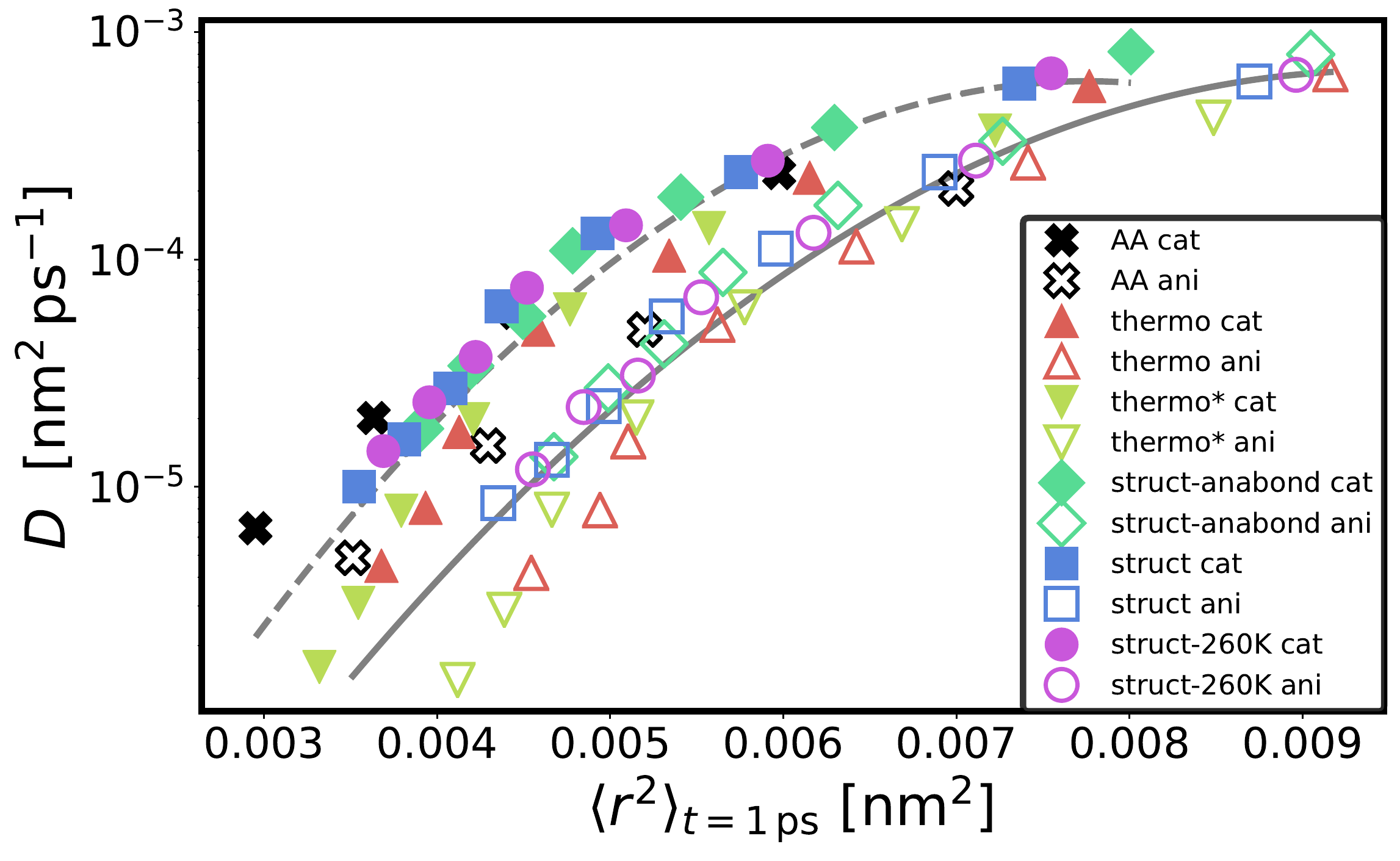}
                \caption{
Diffusion coefficient, $D$, versus the magnitude of the MSD in the plateau region, $r_{\rm p}^2\equiv \langle r^2(t=1\,\rm{ps}) \rangle$. 
Solid (dashed) gray curves are guides for following the trend of anions (cations).
}
                \label{fig:D-v-plat}
        \end{center}
\end{figure}

To ascertain the origin of the observed differences between AA and CG dynamics, we relate long-time and short-time motions for both levels of resolution. 
We anticipate that smoother energy landscapes and higher temperatures result in not only faster diffusive motion but also larger amplitudes of cage rattling motion in the plateau regime. 
Elastic~\cite{Dyre_RMP_06} and elasticity-based~\cite{Schweizer_JPCL_13} theories of glass-forming liquids propose relations between the structural relaxation time and the cage rattling amplitude, which are consistent with simulation results~\cite{Leporini_NAT_PH_08, Douglas_SM_12, Vogel_JPCL_15}.
Fig.~\ref{fig:D-v-plat} presents the diffusion coefficients, $D$, versus the amplitude of the cage rattling motion, which we characterize by the MSD in the plateau regime, $r_{\rm p}^2\equiv \langle r^2(t=1\,\rm{ps}) \rangle$. 
It is apparent that the very different dynamical behaviors of the variety of models characterized above collapse to a similar $D(r_{\rm{p}}^2)$ dependency, where the curves for the cations and anions are somewhat shifted relative to each other. 
Overall, the structure-based models (struct-anabond, struct, struct-260K)  demonstrate a tighter collapse than the thermo and thermo* models, relative to the AA behavior, although these models also show deviations at small $r_{\rm p}^2$ (i.e., lower $T$).
The collapse of CG dynamics implies that a smoothening of the energy landscape leads to a facilitation of the diffusive motion via a softening of the local cages. 
This may provide a promising route for constructing dynamically-consistent CG models, as recently suggested in a study on ILs~\cite{Vogel_CPC_17} and put into practice for liquid ortho-terphenyl~\cite{Xia:2018} and several polymer melts~\cite{Xia:2017,Xia:2019}. 
More specifically, the latter studies argue that a quantitative prediction of long-time dynamics is possible when $r_{\rm p}^2$ is preserved under coarse-graining.
This provides a tractable method for CG parametrizations, since $r_{\rm p}^2$ can be efficiently determined from (relatively) short AA reference simulations.

\section{Conclusions}

This work employed five distinct structure- and/or thermodynamic-based coarse-graining parametrization strategies for modeling the room temperature ionic liquid [C$_4$mim][PF$_6$].
All five models displayed limited transferability in accurately representing structural properties over a range of temperatures, although the absolute error of the structure-based models was much lower overall, by construction.
These structural discrepancies might be systematically addressed via recently proposed approaches for estimating the entropic contribution to the many-body potential of mean force (i.e., the theoretically-ideal CG potential)~\cite{Lebold:2019b, Lebold:2019c, Jin:2019, Jin:2020}.

The focus of our study was to better understand the link between the various parametrization strategies and the dynamical properties of the resulting models.
We quantified the dynamical speedup for each model and for each ionic species via the ratio of diffusion constants, relative to the AA model, $D_{\rm s} = D^{\rm CG} / D^{\rm AA}$.
All five models showed a systematic increase in $D_{\rm s}$, for both cations and anions, with decreasing temperature, indicating dynamical inconsistency relative to the AA model (i.e., a lack of transferability).
At a given temperature, the structure-based models tended to provide a more consistent description of the \emph{relative} anion-to-cation behavior.
Interestingly, the accuracy with which the model described the intramolecular distributions of the cation and also the reference parametrization temperature for the structure-based models appeared to play a limited role in the resulting diffusive properties of the model.
Moreover, our results demonstrate that the \emph{absolute} speedup is independent of the dynamical consistency of the model: in this case the model with the largest speedup provided the most consistent description of anion-cation relative diffusion.

We also investigated the relationship between vibrational and
diffusive motions for each model. Following previous work on theories
of glass-forming liquids, we assessed the relationship between the
structural relaxation time and the cage rattling amplitude via the
diffusion constant and the value of the mean-squared displacement in
the plateau regime, respectively. 
Despite their range of parametrization strategies, we found a collapse of CG model behavior in terms of this relationship (Fig.~\ref{fig:D-v-plat}).
The structure-based models demonstrate a slightly tighter
collapse, relative to the AA behavior, although clear discrepancies
still arise for the lowest temperatures considered. 
While these models reproduce a set of low-dimensional structural distribution functions by construction, the need for iterative refinement relative to the initial (force-matching-based multiscale coarse-graining) model indicates fundamental limitations of the chosen CG interaction functions to accurately describe higher-order structural correlations~\cite{Rudzinski:2014,Rudzinski:2014b,Woerner:2019}.
This then motivates investigations that probe the relationship between improved representations of the many-body potential of mean force and the resulting dynamic consistency of the CG model~\cite{Bereau:2018}.
Overall, the results in this work provide evidence of a clear route to constructing dynamically consistent CG simulation models of RTILs via a temperature-dependent matching of AA vibrational amplitudes, as recently carried out for model glass formers~\cite{Xia:2017,Xia:2018,Xia:2019}.

\section*{Data Availability}

Additional methodological details and results can be found in the Supporting Information of this work.
We have also compiled a repository~\cite{Rudzinski:2021-repo} containing all the models (in GROMACS format) developed in this work (i.e., the thermo*, struct-anabond, struct, and struct-260K models), some parameter files for the calculations performed with the VOTCA and BOCS packages, and also some of the raw data used to construct the manuscript figures.
Additional data can be obtained from the corresponding author upon request.

\section*{Author Contributions}

JFR, KK, TB, and MV designed the work.
JFR, SK, SW, and TP performed calculations for the work.
JFR, SK, KK, TB, and MV analyzed the results.
JFR, SK, KK, TB, and MV wrote the manuscript.

\ack
This
work was supported by the TRR 146 Collaborative Research Center (project A6) of the
Deutsche Forschungsgemeinschaft. 

\section*{References}
\bibliographystyle{unsrt}
\bibliography{biblio,references_MPIP,references_PSU}

\begin{thebibliography}{100}

\bibitem{Atkin_CR_15}
R.~Hayes, G.~G. Warr, and R.~Atkin.
\newblock Structure and nanostructure in ionic liquids.
\newblock {\em Chem. Rev.}, 115:6357--6426, 2015.

\bibitem{Fayer_JCP_18}
A.~Perkin, B.~Kirchner, and M.~D. Fayer.
\newblock Preface: Special topic on chemical physics of ionic liquids.
\newblock {\em J. Chem. Phys.}, 148:193501, 2018.

\bibitem{armand2009ionic}
Michel Armand, Frank Endres, Douglas~R MacFarlane, Hiroyuki Ohno, and Bruno
  Scrosati.
\newblock Ionic-liquid materials for the electrochemical challenges of the
  future.
\newblock {\em Nat. Mater.}, 8(8):621--629, 2009.

\bibitem{Armand_11}
M.~Armand, F.~Endres, D.~R. MacFarlane, H.~Ohno, and B.~Scrosati.
\newblock Ionic-liquid materials for the electrochemical challenges of the
  future.
\newblock In {\em Materials For Sustainable Energy: A Collection of
  Peer-Reviewed Research and Review Articles from Nature Publishing Group},
  pages 129--137. World Scientific, 2011.

\bibitem{Angell_EES_14}
D.~R. MacFarlane, N.~Tachikawa, M.~Forsyth, J.~M. Pringle, P.~C. Howlett, G.~D.
  Elliott, J.~H. Davis, M.~Watanabe, P.~Simon, and C.~A. Angell.
\newblock Energy applications of ionic liquids.
\newblock {\em Energy Environ. Sci.}, 7:232--250, 2014.

\bibitem{Watanabe_CR_17}
M.~Watanabe, M.~L. Thomas, S.~Zhang, K.~Ueno, T.~Yasuda, and K.~Dokko.
\newblock Application of ionic liquids to energy storage and conversion
  materials and devices.
\newblock {\em Chem. Rev.}, 117:7190--7239, 2017.

\bibitem{Russina_JPCB_07}
A.~Triolo, O.~Russina, H.-J. Bleif, and E.~Di~Cola.
\newblock Nanoscale segregation in room temperature ionic liquids.
\newblock {\em J. Phys. Chem. B}, 111:4641--4644, 2007.

\bibitem{Margulis_JPCB_10}
H.~V.~R. Annapureddy, H.~K. Kashyap, P.~M. De~Biase, and C.~J. Margulis.
\newblock What is the origin of the prepeak in the x-ray scattering of
  imidazolium-based room-temperature ionic liquids?
\newblock {\em J. Phys. Chem. B}, 114(50):16838--16846, 2010.

\bibitem{Hardacre_JCP_10}
Ch. Hardacre, J.~D. Holbrey, C.~L. Mullan, T.~G.~A. Youngs, and D.~T. Bowron.
\newblock Small angle neutron scattering from 1-alkyl-3-methylimidazolium
  hexafluorophosphate ionic liquids ([{C$_n$mim][PF$_6$]}, n=4, 6, and 8).
\newblock {\em J. Chem. Phys.}, 133:074510, 2010.

\bibitem{Russina_JPCL_12}
O.~Russina, A.~Triolo, L.~Gontrani, and R.~Caminiti.
\newblock Mesoscopic structural heterogeneities in room-temperature ionic
  liquids.
\newblock {\em J. Phys. Chem. Lett.}, 3:27--33, 2012.

\bibitem{Yamamuro_JCP_15}
Maiko Kofu, Madhusudan Tyagi, Yasuhiro Inamura, Kyoko Miyazaki, and Osamu
  Yamamuro.
\newblock Quasielastic neutron scattering studies on glass-forming ionic
  liquids with imidazolium cations.
\newblock {\em J. Chem. Phys.}, 143(23):234502, 2015.

\bibitem{Rivera_JCP_07}
A.~Rivera, A.~Brodin, A.~Pugachev, and E.~A. R{\"o}ssler.
\newblock Orientational and translational dynamics in room temperature ionic
  liquids.
\newblock {\em J. Chem. Phys.}, 126:114503, 2007.

\bibitem{Kremer_PCCP_09}
J.~R. Sangoro, C.~Iacob, A.~Serghei, C.~Friedrich, and F.~Kremer.
\newblock Universal scaling of charge transport in glass-forming ionic liquids.
\newblock {\em Phys. Chem. Chem. Phys.}, 11:913--916, 2009.

\bibitem{Vogel_JCP_19_MB}
M.~Becher, E.~Steinr\"ucken, and M.~Vogel.
\newblock {\em J. Chem. Phys.}, 151:194503, 2019.

\bibitem{Voth_JPCB_04}
M.~G.~Del Popolo and G.~A. Voth.
\newblock On the structure and dynamics of ionic liquids.
\newblock {\em J. Phys. Chem. B}, 108:1744--1752, 2004.

\bibitem{Padua_JPCB_06}
J.~N.~A. Canongia~Lopes and A.~A.~H. P{\'a}dua.
\newblock Nanostructural organization in ionic liquids.
\newblock {\em J. Phys. Chem. B}, 110:3330--3335, 2006.

\bibitem{wang2007understanding}
Yanting Wang, WEI Jiang, Tianying Yan, and Gregory~A Voth.
\newblock Understanding ionic liquids through atomistic and coarse-grained
  molecular dynamics simulations.
\newblock {\em Acc. Chem. Res.}, 40(11):1193--1199, 2007.

\bibitem{DelleSitte_FD_11}
K.~Wendler, Y.~Y.~Zhao F.~Dommert, R.~Berger, C.~Holm, and L.~{Delle Sitte]}.
\newblock Ionic liquids studied across different scales: A computational
  perspective.
\newblock {\em Faraday Discuss.}, 154:111--132, 2012.

\bibitem{Lopes_JBCS_16}
K.~Shimizu, M.~Tariq, A.~A. Freitas, A.~A.~H. P{\'a}dua, and J.~N.~C. Lopes.
\newblock Self-organization in ionic liquids: From bulk to interfaces and
  films.
\newblock {\em J. Braz. Chem. Soc.}, 27:349--362, 2016.

\bibitem{Maginn2009}
E~J Maginn.
\newblock Molecular simulation of ionic liquids: current status and future
  opportunities.
\newblock {\em J. Phys. Condens. Matter.}, 21(37):373101, August 2009.

\bibitem{Plate_CPC_10}
S.~S. Sarangi, W.~Zhao, F.~M{\"u}ller-Plathe, and S.~Balasubramanian.
\newblock Correlation between dynamic heterogeneity and local structure in a
  room-temperature ionic liquid: A molecular dynamics study of
  [bmim][pf{$_6$}].
\newblock {\em ChemPhysChem.}, 11:2001--2010, 2010.

\bibitem{Wang_JCP_16}
P.~E. Ramirez-Gonzalez, L.~E. Sanchez-Diaz, M.~Medina-Noyola, and Y.~Wang.
\newblock Communication: Probing the existence of partially arrested states in
  ionic liquids.
\newblock {\em J. Chem. Phys.}, 145:191101, 2016.

\bibitem{Kim_JCP_18}
J.~Liu, J.~A.~L. Willcox, and H.~J. Kim.
\newblock Heterogeneous dynamics of ionic liquids: A four-point time
  correlation function approach.
\newblock {\em J. Chem. Phys.}, 148:193830, 2018.

\bibitem{Sulpizi_JCP_18}
K.~Usui, J.~Hunger, M.~Bonn, and M.~Sulpizi.
\newblock Dynamical heterogeneities of rotational motion in room temperature
  ionic liquids evidenced by molecular dynamics simulations.
\newblock {\em J. Chem. Phys.}, 148:193811, 2018.

\bibitem{Smiatek_JCP_18}
A.~Weyman, M.~Bier, Ch. Holm, and J.~Smiatek.
\newblock Microphase separation and the formation of ion conductivity channels
  in poly(ionic liquid)s: A coarse-grained molecular dynamics study.
\newblock {\em J. Chem. Phys.}, 148(19):193824, 2018.

\bibitem{Singer_JPCA_03}
E.~A. Turner, C.~C. Pye, and R.~D. Singer.
\newblock Use of ab initio calculations toward the rational design of room
  temperature ionic liquids.
\newblock {\em J. Phys. Chem. a}, 107:2277--2288, 2003.

\bibitem{Popolo_JPCB_05}
M.~G.~Del Popolo, R.~M. Lynden-Bell, and J.~Kohanoff.
\newblock Ab initio molecular dynamics simulations of a room temperature ionic
  liquid.
\newblock {\em J. Phys. Chem. B}, 109:5895--5902, 2005.

\bibitem{Balasubramanian_CPL_06}
B.~L. Bhargava and S.~Balasubramanian.
\newblock Intermolecular structure and dynamics in an ionic liquid: A
  car–parrinello molecular dynamics simulation study of
  1,3-dimethylimidazolium chloride.
\newblock {\em Chem. Phys. Lett.}, 417:486--491, 2006.

\bibitem{lynden2007simulations}
Ruth~M Lynden-Bell, Mario~G Del~Popolo, Tristan~GA Youngs, Jorge Kohanoff,
  Christof~G Hanke, Jason~B Harper, and Carlos~C Pinilla.
\newblock Simulations of ionic liquids, solutions, and surfaces.
\newblock {\em Acc. Chem. Res.}, 40(11):1138--1145, 2007.

\bibitem{bhargava2008modelling}
BL~Bhargava, Sundaram Balasubramanian, and Michael~L Klein.
\newblock Modelling room temperature ionic liquids.
\newblock {\em Chem. Commun.}, 29:3339--3351, 2008.

\bibitem{Vogel_JPCC_18}
T.~Pal, C.~Beck, D.~Lessnich, and M.~Vogel.
\newblock Effects of silica surfaces on the structure and dynamics of
  room-temperature ionic liquids: A molecular dynamics simulation study.
\newblock {\em J. Phys. Chem. C}, 122:624--634, 2018.

\bibitem{Peter:2009hr}
C.~Peter and K.~Kremer.
\newblock Multiscale simulation of soft matter systems - from the atomistic to
  the coarse-grained level and back.
\newblock {\em Soft Matter}, 5(22):4357--4366, 2009.

\bibitem{Riniker:2012qf}
Sereina Riniker, Jane~R Allison, and Wilfred~F van Gunsteren.
\newblock On developing coarse-grained models for biomolecular simulation: A
  review.
\newblock {\em Phys. Chem. Chem. Phys.}, 14(36):12423--12430, 2012.

\bibitem{noid2013perspective}
W.~G. Noid.
\newblock Perspective: Coarse-grained models for biomolecular systems.
\newblock {\em J. Chem. Phys.}, 139(9):09B201\_1, 2013.

\bibitem{WangVoth2006}
Yanting Wang and Gregory~A. Voth.
\newblock Tail aggregation and domain diffusion in ionic liquids.
\newblock {\em J. Phys. Chem. B}, 110(37):18601--18608, September 2006.

\bibitem{Bhargava2007}
B.~Lokegowda Bhargava, Russell Devane, Michael~L. Klein, and Sundaram
  Balasubramanian.
\newblock Nanoscale organization in room temperature ionic liquids: a coarse
  grained molecular dynamics simulation study.
\newblock {\em Soft Matter}, 3(11):1395, 2007.

\bibitem{KarimiVarzaneh2010}
Hossein~Ali Karimi-Varzaneh, Florian M\"{u}ller-Plathe, Sundaram
  Balasubramanian, and Paola Carbone.
\newblock Studying long-time dynamics of imidazolium-based ionic liquids with a
  systematically coarse-grained model.
\newblock {\em Phys. Chem. Chem. Phys.}, 12(18):4714, 2010.

\bibitem{Tokuda2004}
Hiroyuki Tokuda, Kikuko Hayamizu, Kunikazu Ishii, Md. Abu Bin~Hasan Susan, and
  Masayoshi Watanabe.
\newblock Physicochemical properties and structures of room temperature ionic
  liquids. 1. variation of anionic species.
\newblock {\em J. Phys. Chem. B}, 108(42):16593--16600, October 2004.

\bibitem{Deichmann2019}
Gregor Deichmann and Nico F.~A. van~der Vegt.
\newblock Conditional reversible work coarse-grained models with explicit
  electrostatics{\textemdash}an application to butylmethylimidazolium ionic
  liquids.
\newblock {\em J. Chem. Theory Comput.}, 15(2):1187--1198, January 2019.

\bibitem{Laaksonen_PCCP_13}
Y.-L. Wang, A.~Lyubartsev, Z.-Y. Lu, and A.~Laaksonen.
\newblock Multiscale coarse-grained simulations of ionic liquids: Comparison of
  three approaches to derive effective potentials.
\newblock {\em Phys. Chem. Chem. Phys.}, 15:7701--7712, 2013.

\bibitem{Aluru_JCTC_18}
A.~Moradzadeh, M.~H. Motevaselian, S.~Y. Mashayak, and N.~R. Aluru.
\newblock Coarse-grained force fields for imidazolium-based ionic liquids.
\newblock {\em J. Chem. Theory Comput.}, 14:3252--3261, 2018.

\bibitem{Ruza:2020}
Jurgis Ruza, Wujie Wang, Daniel Schwalbe-Koda, Simon Axelrod, William~H.
  Harris, and Rafael G\'{o}mez-Bombarelli.
\newblock Temperature-transferable coarse-graining of ionic liquids with dual
  graph convolutional neural networks.
\newblock {\em J. Chem. Phys.}, 153(16):164501, 2020.

\bibitem{Mukherjee:2017}
Biswaroop Mukherjee, Christine Peter, and Kurt Kremer.
\newblock {Single molecule translocation in smectics illustrates the challenge
  for time-mapping in simulations on multiple scales}.
\newblock {\em J. Chem. Phys.}, 147(11):114501, 2017.

\bibitem{Rudzinski2019}
Joseph~F. Rudzinski.
\newblock Recent progress towards chemically-specific coarse-grained simulation
  models with consistent dynamical properties.
\newblock {\em Computation}, 7(3):42, August 2019.

\bibitem{Harmandaris:2009oc}
V.~A. Harmandaris and K.~Kremer.
\newblock Predicting polymer dynamics at multiple length and time scales.
\newblock {\em Soft Matter}, 5(20):3920--3926, 2009.

\bibitem{Salerno:2016}
K.M. Salerno, A.~Agrawal, D.~Perahia, and G.S. Grest.
\newblock Resolving dynamic properties of polymers through coarse-grained
  computational studies.
\newblock {\em Phys. Rev. Lett.}, {116}:058302, {2016}.

\bibitem{Rudzinski:2016}
Joseph~F. Rudzinski, Kurt Kremer, and Tristan Bereau.
\newblock Communication: Consistent interpretation of molecular simulation
  kinetics using markov state models biased with external information.
\newblock {\em J. Chem. Phys.}, 144(5):051102, 2016.

\bibitem{Rudzinski:2016b}
Joseph~F. Rudzinski and Tristan Bereau.
\newblock {Concurrent parametrization against static and kinetic information
  leads to more robust coarse-grained force fields}.
\newblock {\em Eur. Phys. J. Special Topics}, {225}({8-9}):1373--1389, {2016}.

\bibitem{Bereau:2018}
Tristan Bereau and Joseph~F Rudzinski.
\newblock Accurate structure-based coarse-graining leads to consistent
  barrier-crossing dynamics.
\newblock {\em Phys. Rev. Lett.}, 121:256002, 2018.

\bibitem{Larini:2010dq}
L.~Larini, L.~Lu, and G.~A. Voth.
\newblock The multiscale coarse-graining method. vi. implementation of
  three-body coarse-grained potentials.
\newblock {\em J. Chem. Phys.}, 132(16):164107, 2010.

\bibitem{Das:2012c}
Avisek Das and Hans~C. Andersen.
\newblock {The Multiscale Coarse-Graining Method. IX. A General Method for
  Construction of Three Body Coarse-Grained Force Fields}.
\newblock {\em J. Chem. Phys.}, {136}({19}):{194114}, {2012}.

\bibitem{Lindsey:2017}
Rebecca~K. Lindsey, Laurence~E. Fried, and Nir Goldman.
\newblock {ChIMES: A Force Matched Potential with Explicit Three-Body
  Interactions for Molten Carbon}.
\newblock {\em J. Chem. Theor. Comp.}, {13}({12}):6222--6229, {2017}.

\bibitem{Scherer:2018}
Christoph Scherer and Denis Andrienko.
\newblock Understanding three-body contributions to coarse-grained force
  fields.
\newblock {\em Phys. Chem. Chem. Phys.}, 20:22387--22394, 2018.

\bibitem{John:2017}
Sebastian~T John and G{\'a}bor Cs{\'a}nyi.
\newblock Many-body coarse-grained interactions using gaussian approximation
  potentials.
\newblock {\em J. Phys. Chem. B}, 121(48):10934--10949, 2017.

\bibitem{Zhang:2018}
Linfeng Zhang, Jiequn Han, Han Wang, Roberto Car, and Weinan E.
\newblock {DeePCG: Constructing coarse-grained models via deep neural
  networks}.
\newblock {\em J. Chem. Phys.}, {149}({3}):{034101}, {2018}.

\bibitem{Wang:2019}
Jiang Wang, Simon Olsson, Christoph Wehmeyer, Adria Perez, Nicholas~E. Charron,
  Gianni de~Fabritiis, Frank Noe, and Cecilia Clementi.
\newblock Machine learning of coarse-grained molecular dynamics force fields.
\newblock {\em ACS Cent. Sci.}, {5}({5}):755--767, 2019.

\bibitem{Chan:2019}
Henry Chan, Mathew~J. Cherukara, Badri Narayanan, Troy~D. Loeffler, Chris
  Benmore, Stephen~K. Gray, and Subramanian K. R.~S. Sankaranarayanan.
\newblock {Machine learning coarse grained models for water}.
\newblock {\em Nat. Commun.}, {10}:{379}, 2019.

\bibitem{Scherer:2020}
Christoph Scherer, Rene Scheid, Denis Andrienko, and Tristan Bereau.
\newblock Kernel-based machine learning for efficient simulations of molecular
  liquids.
\newblock {\em J. Chem. Theor. Comp.}, {16}({5}):3194--3204, 2020.

\bibitem{Shahidi:2020}
Nobahar Shahidi, Antonis Chazirakis, Vagelis Harmandaris, and Manolis
  Doxastakis.
\newblock {Coarse-graining of polyisoprene melts using inverse Monte Carlo and
  local density potentials}.
\newblock {\em J. Chem. Phys.}, {152}({12}):124902, 2020.

\bibitem{Zuo:2020}
Yunxing Zuo, Chi Chen, Xiangguo Li, Zhi Deng, Yiming Chen, J\"{o}rg Behler,
  G{\'{a}}bor Cs{\'{a}}nyi, Alexander~V. Shapeev, Aidan~P. Thompson,
  Mitchell~A. Wood, and Shyue~Ping Ong.
\newblock Performance and cost assessment of machine learning interatomic
  potentials.
\newblock {\em J. Phys. Chem. A}, 124(4):731--745, 2020.

\bibitem{DeLyser:2017}
Michael~R. DeLyser and William~G. Noid.
\newblock {Extending pressure-matching to inhomogeneous systems via
  local-density potentials}.
\newblock {\em J. Chem. Phys.}, {147}({13}):134111, 2017.

\bibitem{Sanyal:2018}
Tanmoy Sanyal and M~Scott Shell.
\newblock Transferable coarse-grained models of liquid--liquid equilibrium
  using local density potentials optimized with the relative entropy.
\newblock {\em J. Phys. Chem. B}, 122(21):5678--5693, 2018.

\bibitem{Jin:2018}
Jaehyeok Jin, Yining Han, and Gregory~A. Voth.
\newblock Ultra-coarse-grained liquid state models with implicit hydrogen
  bonding.
\newblock {\em J. Chem. Theor. Comp.}, {14}({12}):6159--6174, 2018.

\bibitem{DeLyser:2019}
Michael~R. DeLyser and W.~G. Noid.
\newblock {Analysis of local density potentials}.
\newblock {\em J. Chem. Phys.}, {151}({22}):224106, 2019.

\bibitem{Davtyan:2014}
Aram Davtyan, James~F Dama, Anton~V Sinitskiy, and Gregory~A Voth.
\newblock The theory of ultra-coarse-graining. 2. numerical implementation.
\newblock {\em J. Chem. Theor. Comp.}, 10(12):5265--5275, 2014.

\bibitem{Katkar:2018}
Harshwardhan~H. Katkar, Aram Davtyan, Aleksander E.~P. Durumeric, Glen~M.
  Hocky, Anthony~C. Schramm, Enrique~M. De~La~Cruz, and Gregory~A. Voth.
\newblock Insights into the cooperative nature of atp hydrolysis in actin
  filaments.
\newblock {\em Biophys. J.}, {115}({8}):1589--1602, 2018.

\bibitem{Dama:2017}
James~F Dama, Jaehyeok Jin, and Gregory~A Voth.
\newblock The theory of ultra-coarse-graining. 3. coarse-grained sites with
  rapid local equilibrium of internal states.
\newblock {\em J. Chem. Theor. Comp.}, 13(3):1010--1022, 2017.

\bibitem{Rudzinski:2020}
Joseph~F. Rudzinski and Tristan Bereau.
\newblock Coarse-grained conformational surface hopping: Methodology and
  transferability.
\newblock {\em J. Chem. Phys.}, 153:214110, 2020.

\bibitem{Vogel_CPC_17}
T.~Pal and M.~Vogel.
\newblock Role of dynamic heterogeneities in ionic liquids: Insights from
  all-atom and coarse-grained molecular dynamics simulation studies.
\newblock {\em ChemPhysChem.}, 18:2233--2242, 2017.

\bibitem{Vogel_JCP_19_TP}
T.~Pal and M.~Vogel.
\newblock On the relevance of electrostatic interactions for the structural
  relaxation of ionic liquids: A molecular dynamics simulation study.
\newblock {\em J. Chem. Phys.}, 150:124501, 2019.

\bibitem{Vogel_JCP_15_PH}
P.~Henritzi, A.~Bormuth, F.~Klameth, and M.~Vogel.
\newblock A molecular dynamics simulations study on the relations between
  dynamical heterogeneity, structural relaxation, and self-diffusion in viscous
  liquids.
\newblock {\em J. Chem. Phys.}, 143:164502, 2015.

\bibitem{Xia:2017}
Wenjie Xia, Jake Song, Cheol Jeong, David~D. Hsu, Frederick~R. Phelan, Jr.,
  Jack~F. Douglas, and Sinan Keten.
\newblock {Energy-Renormalization for Achieving Temperature Transferable
  Coarse-Graining of Polymer Dynamics}.
\newblock {\em Macromolecules}, {50}({21}):8787--8796, {2017}.

\bibitem{Xia:2018}
Wenjie Xia, Jake Song, Nitin~K. Hansoge, Frederick~R. Phelan, Sinan Keten, and
  Jack~F. Douglas.
\newblock Energy renormalization for coarse-graining the dynamics of a model
  glass-forming liquid.
\newblock {\em J. Phys. Chem. B}, 122(6):2040--2045, 2018.

\bibitem{Song:2018}
Jake Song, David~D. Hsu, Kenneth~R. Shull, Frederick~R. Phelan, Jack~F.
  Douglas, Wenjie Xia, and Sinan Keten.
\newblock Energy renormalization method for the coarse-graining of polymer
  viscoelasticity.
\newblock {\em Macromolecules}, 51(10):3818--3827, 2018.

\bibitem{Xia:2019}
Wenjie Xia, Nitin~K. Hansoge, Wen-Sheng Xu, Frederick~R. Phelan, Jr., Sinan
  Keten, and Jack~F. Douglas.
\newblock {Energy renormalization for coarse-graining polymers having different
  segmental structures}.
\newblock {\em Sci. Adv.}, {5}({4}):eaav4683, {2019}.

\bibitem{Bhargava2007b}
B.~L. Bhargava and S.~Balasubramanian.
\newblock Refined potential model for atomistic simulations of ionic liquid
  [bmim][{PF}6].
\newblock {\em J. Chem. Phys.}, 127(11):114510, September 2007.

\bibitem{GROMACS5}
M.~J. Abraham, T.~Murtola, R.~Schulz, S.~Pall, J.C. Smith, B.~Hess, and
  E.~Lindahl.
\newblock Gromacs: High performance molecular simulations through multi-level
  parallelism from laptops to supercomputers.
\newblock {\em SoftwareX}, 1-2:19--25, 2015.

\bibitem{Darden:1993}
Tom Darden, Darrin York, and Lee Pedersen.
\newblock Particle mesh ewald: {A}n { N log(N)} method for ewald sums in large
  systems.
\newblock {\em J. Chem. Phys.}, 99({12}):8345--8348, 1993.

\bibitem{NH1}
S.~Nos{\'e}.
\newblock A unified formulation of the constant temperature molecular dynamics
  methods.
\newblock {\em J. Chem. Phys}, 81:511, 1984.

\bibitem{NH2}
W.~G. Hoover.
\newblock Canonical dynamics: Equilibrium phase-space distributions.
\newblock {\em Phys. Rev. A}, 31:1695--1697, 1985.

\bibitem{Parrinello:1982}
M.~Parrinello and A.~Rahman.
\newblock Strain fluctuations and elastic constants.
\newblock {\em J. Chem. Phys.}, 76({5}):2662--2666, 1982.

\bibitem{Soper:1996ly}
A.~K. Soper.
\newblock Empirical potential monte carlo simulation of fluid structure.
\newblock {\em Chem. Phys.}, 202(2-3):295--306, 1996.

\bibitem{Reith:2001}
Dirk Reith, Hendrik Meyer, and Florian M\"uller-Plathe.
\newblock Mapping atomistic to coarse-grained polymer models using automatic
  simplex optimization to fit structural properties.
\newblock {\em Macromolecules}, 34:2335--45, 2001.

\bibitem{Mullinax:2009b}
J.~W. Mullinax and W.~G. Noid.
\newblock Generalized {Yvon-Born-Green} theory for molecular systems.
\newblock {\em Phys. Rev. Lett.}, 103(19):198104, 2009.

\bibitem{Mullinax:2010}
J.~W. Mullinax and W.~G. Noid.
\newblock A generalized {Yvon-Born-Green} theory for determining coarse-grained
  interaction potentials.
\newblock {\em J. Phys. Chem. C}, 114({12}):5661--5674, 2010.

\bibitem{Rudzinski:2012vn}
Joseph~F Rudzinski and William~G Noid.
\newblock The role of many-body correlations in determining potentials for
  coarse-grained models of equilibrium structure.
\newblock {\em J. Phys. Chem. B}, 116(29):8621--35, 2012.

\bibitem{Rudzinski:2015}
Joseph~F. Rudzinski and Will~G. Noid.
\newblock {A generalized-Yvon-Born-Green method for coarse-grained modeling
  Advances, challenges, and insight}.
\newblock {\em Eur. Phys. J. Special Topics}, {224}({12}):2193--2216, {2015}.

\bibitem{Noid:2008a}
W.~G. Noid, J.-W. Chu, G.~S. Ayton, V.~Krishna, S.~Izvekov, G.~A. Voth, A.~Das,
  and H.~C. Andersen.
\newblock The multiscale coarse-graining method. {I.} {A} rigorous bridge
  between atomistic and coarse-grained models.
\newblock {\em J. Chem. Phys.}, 128({24}):244114, 2008.

\bibitem{Rudzinski:2014}
Joseph~F. Rudzinski and W.~G. Noid.
\newblock {Investigation of Coarse-Grained Mappings via an Iterative
  Generalized Yvon-Born-Green Method}.
\newblock {\em J. Phys. Chem. B}, 118({28}):8295--8312, 2014.

\bibitem{Cho:2009ve}
H.~M. Cho and J.~W. Chu.
\newblock Inversion of radial distribution functions to pair forces by solving
  the yvon-born-green equation iteratively.
\newblock {\em J. Chem. Phys.}, 131(13):134107, 2009.

\bibitem{Lu:2013uq}
Lanyuan Lu, James~F. Dama, and Gregory~A. Voth.
\newblock Fitting coarse-grained distribution functions through an iterative
  force-matching method.
\newblock {\em J. Chem. Phys.}, 139(12):121906, 2013.

\bibitem{Rudzinski:2014b}
Joseph~F. Rudzinski and W.~G. Noid.
\newblock {Bottom-Up Coarse-Graining of Peptide Ensembles and Helix-Coil
  Transitions}.
\newblock {\em J. Chem. Theor. Comp.}, {11}({3}):1278--1291, 2015.

\bibitem{Rudzinski:2021-repo}
\texttt{Dynamical properties across different coarse-grained models for ionic
  liquids} repository.
\newblock
  \url{https://gitlab.mpcdf.mpg.de/jrudz/models_for_dynamical_properties_across_different_coarse-grained_models_for_ionic_liquids.git}.

\bibitem{Ruhle:2009wx}
V.~R{\"u}hle, C.~Junghans, A.~Lukyanov, K.~Kremer, and D.~Andrienko.
\newblock Versatile object-oriented toolkit for coarse-graining applications.
\newblock {\em J. Chem. Theor. Comp.}, 5(12):3211--3223, 2009.

\bibitem{Dunn:2018}
Nicholas J.~H. Dunn, Kathryn~M. Lebold, Michael~R. DeLyser, Joseph~F.
  Rudzinski, and W.G. Noid.
\newblock Bocs: Bottom-up open-source coarse-graining software.
\newblock {\em J. Phys. Chem. B}, 122(13):3363--3377, 2018.

\bibitem{Mullinax:2010a}
J.~W. Mullinax and W.~G. Noid.
\newblock Reference state for the generalized {Yvon-Born-Green} theory:
  Application for coarse-grained model of hydrophobic hydration.
\newblock {\em J. Chem. Phys.}, 133({12}):124107, 2010.

\bibitem{Hess:2008}
Berk Hess, Carsten Kutzner, David van~der Spoel, and Erik Lindahl.
\newblock Gromacs 4: Algorithms for highly efficient, load-balanced, and
  scalable molecular simulation.
\newblock {\em J. Chem. Theor. Comp.}, 4(3):435--447, 2008.

\bibitem{Johnson2007}
Margaret~E. Johnson, Teresa Head-Gordon, and Ard~A. Louis.
\newblock Representability problems for coarse-grained water potentials.
\newblock {\em The Journal of Chemical Physics}, 126(14):144509, April 2007.

\bibitem{Wang:2009ol}
H.~Wang, C.~Junghans, and K.~Kremer.
\newblock Comparative atomistic and coarse-grained study of water: What do we
  lose by coarse-graining?
\newblock {\em Eur. Phys. J. E}, 28(2):221--229, 2009.

\bibitem{Guenza:2015}
M.~Guenza.
\newblock {Thermodynamic consistency and other challenges in coarse-graining
  models}.
\newblock {\em Eur. Phys. J. Special Topics}, {224}({12}):2177--2191, 2015.

\bibitem{Dunn:2016b}
Nicholas J.~H. Dunn, Thomas~T. Foley, and William~G. Noid.
\newblock {Van der Waals Perspective on Coarse-Graining: Progress toward
  Solving Representability and Transferability Problems}.
\newblock {\em Acc. Chem. Res.}, {49}({12}):2832--2840, {2016}.

\bibitem{Robbins:1988}
MO~Robbins, K~Kremer, and GS~Grest.
\newblock {Phase-Diagram and Dyanmics of Yukawa Systems}.
\newblock {\em J. Chem. Phys.}, {88}({5}):3286--3312, 1988.

\bibitem{Dyre_RMP_06}
J.~C. Dyre.
\newblock Colloquium: The glass transition and elastic models of glass-forming
  liquids.
\newblock {\em Rev. Mod. Phys.}, 78:953--972, 2006.

\bibitem{Schweizer_JPCL_13}
S.~Mirigian and K.~S. Schweizer.
\newblock Unified theory of activated relaxation in liquids over 14 decades in
  time.
\newblock {\em J. Phys. Chem. Lett.}, 4:3648--3653, 2015.

\bibitem{Leporini_NAT_PH_08}
L.~Larini, A.~Ottochian, C.~De Michele, and D.~Leporini.
\newblock Universal scaling between structural relaxation and vibrational
  dynamics in glass-forming liquids and polymers.
\newblock {\em Nat. Phys.}, 4:42--45, 2008.

\bibitem{Douglas_SM_12}
D.~S. Simmons, M.~T. Cicerone, Q.~Zhong, M.~Tyagi, and J.~F. Douglas.
\newblock Generalized localization model of relaxation in glass-forming
  liquids.
\newblock {\em Soft Matter}, 8:11455--11461, 2012.

\bibitem{Vogel_JPCL_15}
F.~Klameth and M.~Vogel.
\newblock Slow water dynamics near a glass transition or a solid interfacs: A
  common rationale.
\newblock {\em J. Phys. Chem. Lett.}, 6:4385--4389, 2015.

\bibitem{Lebold:2019b}
Kathryn~M. Lebold and W.~G. Noid.
\newblock {Dual approach for effective potentials that accurately model
  structure and energetics}.
\newblock {\em J. Chem. Phys.}, {150}({23}):234107, {2019}.

\bibitem{Lebold:2019c}
Kathryn~M. Lebold and W.~G. Noid.
\newblock {Dual-potential approach for coarse-grained implicit solvent models
  with accurate, internally consistent energetics and predictive
  transferability}.
\newblock {\em J. Chem. Phys.}, {151}({16}):164113, {2019}.

\bibitem{Jin:2019}
Jaehyeok Jin, Alexander~J. Pak, and Gregory~A. Voth.
\newblock {Understanding Missing Entropy in Coarse-Grained Systems: Addressing
  Issues of Representability and Transferability}.
\newblock {\em J. Phys. Chem. Lett.}, {10}({16}):4549--4557, {2019}.

\bibitem{Jin:2020}
Jaehyeok Jin, Alvin Yu, and Gregory~A. Voth.
\newblock Temperature and phase transferable bottom-up coarse-grained models.
\newblock {\em J. Chem. Theor. Comp.}, 16(11):6823--6842.

\bibitem{Woerner:2019}
Svenja~J. Woerner, Tristan Bereau, Kurt Kremer, and Joseph~F. Rudzinski.
\newblock {Direct route to reproducing pair distribution functions with
  coarse-grained models via transformed atomistic cross correlations}.
\newblock {\em J. Chem. Phys.}, {151}({24}):244110, 2019.

\end{thebibliography}

\end{document}